\documentclass[journal]{IEEEtran}
\usepackage{amsmath}
\usepackage{cite}
\usepackage{bm}
\usepackage{amsfonts}
\usepackage{graphicx}
\usepackage{float}
\usepackage[justification=centering]{caption}
\usepackage{multirow}
\usepackage{diagbox}
\usepackage{amssymb}
\usepackage{bbding}
\usepackage{tabularx}
\usepackage{stfloats}
\usepackage{enumerate}
\usepackage{enumitem}

\ifCLASSINFOpdf
\else
\fi
\setlist[itemize,1]{label=$\bullet$}

\begin{document}
%
\title{Proxy Re-encryption based Fair Trade Protocol for Digital Goods Transactions on Smart Contracts}
%
%
%

\author{Peng~Zhang,
        Jiaquan~Wei,
        Yuhong~Liu,
        Hongwei~Liu
\thanks{Y. Liu is the corresponding author.}
\thanks{This work was in part supported by the National Natural Science Foundation of China under Grant 61702342 and Grant 61872243.}
\thanks{P. Zhang and J. Wei are with the College
of Electronics and Information Engineering, Shenzhen University, Shenzhen 518060, China (e-mail: zhangp@szu.edu.cn; 798906320@qq.com ).}
\thanks{Y. Liu is with the Department of Computer Engineering, Santa Clara University, Santa Clara 95053, USA (e-mail: yhliu@scu.edu).}
\thanks{H. Liu is with the Shenzhen Technology University, Shenzhen 518118, China (e-mail: liuhongwei@sztu.edu.cn).}
}

%
%

\markboth{Journal of \LaTeX\ Class Files,~Vol.~*, No.~*, May~2021}%
{Shell \MakeLowercase{\textit{et al.}}: Bare Demo of IEEEtran.cls for IEEE Journals}
%



\maketitle

\begin{abstract}
With the massive amount of digital data generated everyday, transactions of digital goods become a trend, and the fairness of such transactions has drawn increasing attentions in recent years.
Fairness is defined as that both of the seller and the buyer get what they want, or neither, which is one of the most essential requirements for transactions.
Researches show that it is impossible to design a fair protocol only with a seller and a buyer.
Thus current fair protocols generally rely on a trusted third-party (TTP) to trade.
However, the fairness is based on the TTP's behaviors and the two parties' trust in the TTP.
With the emergence of Blockchain and smart contract, its decentralization and transparency make it a very good candidate to replace the TTP.
So in this work, we attempt to design a secure and fair protocol for digital goods transactions based on smart contracts.
To ensure security of the digital goods, we proposed an advanced passive proxy re-encryption (PRE) scheme, which makes smart contracts transfer decryption right to a buyer after receiving his/her payment.
Furthermore, based on smart contracts and the proposed passive PRE scheme, a fair protocol for digital goods transactions is proposed, whose fairness is guaranteed by the arbitration protocol.
In addition, the proposed protocol supports ciphertext publicity and repeatable sale, which achieves fewer interaction times.
Comprehensive experiment results validate the feasibility and effectiveness of the proposed protocol.

\end{abstract}

%
\IEEEpeerreviewmaketitle

\section{Introduction}
%
%
%
%
\IEEEPARstart{D}{igital} goods, also named electronic goods or e-goods, refers to any products or services that are stored, delivered, and used in their digital forms. Examples of digital goods include data, audio or video files, games, digital currency, software, design, and computing services. Digital goods takes a large and rapid growing share of our economy \cite{brynjolfsson2019should} and is therefore attracting increasing attention on its fairness.
In a typical digital goods transaction between two parties (i.e. a buyer and a seller), the buyer has to provide payment first and will only receive the released digital goods when the seller successfully confirms the payment.
The result of the transaction completely depends on the seller's behavior, which is obviously unfair.

Therefore, one critical challenge in digital goods transactions is how to guarantee the fairness, that is, both parties get what they want, or neither.
Fairness is critical to ensure that any honest parties are not at a disadvantage.
Besides fairness, a fair trade protocol is expected to also possess other desirable properties, including privacy, anonymity, and non-repudiability.
Privacy means that the context of the digital goods cannot be obtained by unauthorized users.
Anonymity emphasises that the identities of both parties are not leaked.
Non-repudiability means neither party repudiates its own behaviors.

Focusing on fairness, most existing solutions introduce a trusted third-party (TTP) to participate in digital goods transactions \cite{DBLP:conf/sp/ZhouG96}, who securely hosts the digital goods on its own platform.
When TTP receives payment from a buyer, the digital goods will be released to the buyer, and the payment is transferred to the seller only after the transaction completes.
During this process, TTP undertakes the functions of pre-guarantee and dispute resolution to ensure that the seller gets paid if and only if the buyer gets the digital goods, so that the fairness of transactions is guaranteed. However, there are some remaining issues in such TTP-based digital goods transactions.
(1) The transaction is based on the assumption that both the buyer and the seller fully trust the TTP.
(2) The fairness of transactions depends on the behavior of TTP.
(3) Digital goods is directly exposed to TTP, lacking of data privacy and copyright protection.
(4) TTP may extract transaction fees, resulting in high transaction costs of digital goods.

The emergence of Blockchain \cite{nakamoto2008bitcoin} technologies provides another possibility for digital goods transactions.
The decentralization and transparency of Blockchain can effectively prevent it from taking sides in transactions, which makes it a potential fair transaction platform trusted by both the seller and the buyer sides.
However, some challenges also need to be taken into account.
Firstly, as the size of digital goods involved in transactions may be large, it is difficult to store it directly on Blockchain.
Secondly, information stored on Blockchain is by default open to the public, raising privacy concerns. Encrypting such data, however, brings new challenges to the fairness, as after encryption, it is difficult to verify whether the original data is identical with the claimed digital goods.

In this work, we attempt to design a secure and fair trade protocol of digital goods based on Blockchain by addressing the above challenges.
In particular, IPFS (InterPlanetary File System) \cite{DBLP:journals/corr/Benet14} is adopted to store digital goods data outside the Blockchain, which addresses the limited data storage capacity of Blockchain.
Furthermore, data privacy is protected through encryption technologies, so that any unpaid parties cannot access the digital goods.
More importantly, a passive proxy re-encryption (PRE) is proposed to verify data consistency  without decrypting it.
In particular, smart contracts \cite{buterin2014next} are chosen to take on the role of a proxy, checking data consistency and transferring the decryption right.
Specifically, smart contracts can automatically execute cipertext transform after payment to ensure that the buyer gets the digital goods and the seller receives the payment.
Our major contributions are summarized as follows.

\begin{itemize}
  \item We define and propose an advanced passive PRE. Existing literature mainly focuses on active PRE, where the delegator decides whom he/she delegates the decryption right to. These active PRE schemes also assume the honesty of the delegator. Such schemes, however, cannot be applied to the digital goods transactions since (1) the seller may not be honest; and (2) the seller cannot know who will buy his/her goods until the buyer makes a request. A passive PRE is designed in this work to facilitate digital goods transactions, where the delegation will only be enabled when a seller (as a delegator) receives a request from a buyer (as a delegatee).
  \item By connecting the re-encryption key with the original ciphtertext, the proposed passive PRE scheme is backward-secure when collusions between the proxy and the delegatee exist. Although some collusion-secure schemes are proposed to avoid the leakage of the delegator's private key, they cannot effectively prevent the illegal decryption of the other ciphertexts under the same private key. Thus, backward security is important. That means the unsold ciphertexts under the same private key are secure even if the proxy and the delegatee collude. This is a critical property to ensure fair digital goods transactions.
  \item We propose a fair trade protocol for digital goods, which adopts smart contracts as the proxy of the proposed passive PRE. The transaction fairness is guaranteed by the decentralized characteristics of smart contracts as well as the proposed re-encryption algorithm and arbitration procedure.
\end{itemize}

\section{Related work}
\subsection{TTP-based fair protocols}
The TTP-based fair protocols can be divided into two categories based on whether the TTP involved is online or offline.
The online TTP acts as an intermediary providing network services to all parties to ensure  the fairness of the exchange \cite{DBLP:conf/sp/ZhouG96}\cite{DBLP:conf/ccs/FranklinR97}.
However the online TTP is always involved in the protocol even if all parties are honest, easily becoming the bottleneck of computations and communications.
On the other hand, an offline TTP does not participate in the protocol if all parties act honestly, unless a dispute occurs \cite{DBLP:conf/csfw/ZhouG97}\cite{DBLP:conf/sp/BaoDM98}.
Therefore offline TTP-based protocols are also called an optimistic protocol \cite{DBLP:journals/jsac/AsokanSW00}.
Furthermore, when a party cannot leave the protocol with a even small advantage over the other party, we call it a strong fair exchange \cite{DBLP:journals/sigecom/RayR02}.
A weak fair exchange is where a misbehaving party can be identified and penalized in case of a dispute  \cite{DBLP:journals/sigecom/RayR02}.
In this paper, the proposed scheme aims to achieve strong fair exchange.

\subsection{Blockchain-based fair protocols}
Recently, Blockchain is introduced to fair exchange protocols to achieve various goals.
In \cite{DBLP:conf/icebe/JayasingheMM14}\cite{DBLP:conf/crypto/BentovK14}, Bitcoin is regraded as a means of implementing a penalty mechanism.
Following these works, similar approaches to implement penalty based incentives through smart contracts have been proposed in \cite{ZhaoYLHD19}\cite{XiongX19}.
Zhao et al. \cite{ZhaoYLHD19} integrated ring signature, double-authentication-preventing signature and similarity learning to guarantee the availability of trading data, privacy of data providers, and fairness between data providers and data consumers.
Xiong et al. \cite{XiongX19} proposed a smart contract-based data trading model solution using Blockchain and machine learning.
Generally speaking, these protocols all require the participation of a TTP. When a dispute occurs, TTP must intervene to resolve the dispute.

Different from these above studies, some literature achieves fair trade without relying on the use of a TTP, and Blockchain is actively involved in the executement of protocols.
Dziembowski et al. \cite{DziembowskiEF18} proposed an efficient protocol for fair trade of digital goods using smart contracts without a trusted third-party.
In their study, smart contracts took the role of an judge that resolved dispute. However, in each transaction, the seller was required to disclose the symmetric key to the smart contracts, which may lead to potential privacy breach.
Guan et al. \cite{GuanSW18} divided the data plaintext into several blocks, encrypted them separately, and uploaded all the ciphertext blocks and Merkle to the smart contracts. However, it was only a semi-automatic system, which required the seller and buyer to exchange an encryption key and a hash value off the Blockchain.
Asgaonkar \cite{AsgaonkarK19} proposed a dual-deposit escrow smart contract for provably cheat-proof delivery and payment for a digital good.

Different from these studies, we aim to achieve transaction fairness in a fully automatic way without revealing encryption keys to smart contracts. To this end, a passive proxy re-encryption scheme is proposed.


\subsection{Proxy re-encryption}
In 1998, Blaze el al. \cite{BlazeBS98} proposed the concept of proxy re-encryption (PRE), where a semi-trusted proxy can transform a ciphertext under Alice's public key into another ciphertext under Bob's public key, with the constraint that Bob can decrypt it with his own private key, but the proxy cannot learn anything about the plaintext.
PRE schemes are divided into two categories: bidirectional and unidirectional, depending on whether the proxy can convert the ciphertext in both directions or one direction only.
In the scenario of a fair digital goods transaction, unidirectional PRE is required since it can prevent the proxy from performing ciphertext conversion in the other direction without permission.

Ateniese el al. \cite{AtenieseFGH05} proposed the first unidirectional PRE scheme using bilinear pairing, which however cannot resist chosen-ciphertext attack (CCA).
Libert el al. \cite{LibertV08} proposed a unidirectional PRE scheme using bilinear pairing, but it only met replayable chosen-ciphertext security.
Wang el al. \cite{WangC13} proposed a unidirectional CCA-secure PRE scheme using bilinear pairing.
As the computational cost of bilinear pairing is much higher than that of modular exponentiation, which may lead to significantly higher economic costs on smart contracts, we are inclined to the PRE schemes without bilinear pairing.

Weng et al. \cite{DengWLC08} and Deng et al. \cite{WengDLC10} constructed PRE schemes without bilinear pairing, which meet CCA security. However, they are bidirectional PRE schemes.
Shao et al. \cite{ShaoC09} proposed a unidirectional PRE schemes without bilinear pairing.
However, it failed to satisfy CCA security when facing specific attacks \cite{ChowWYD10}. Although a more efficient unidirectional PRE scheme without bilinear pairing was proposed in \cite{ChowWYD10}, an important flaw in the security proof of PRE scheme was identified by \cite{SelviPR17}, and an anti-collusion, unidirectional PRE scheme without bilinear pairing was proposed, which satisfied CCA security under the random oracle model.

All the PRE schemes discussed above can be considered as active PRE, where the delegator needs to decide whom the decryption right should be delegated to. However, in a digital goods transaction, the seller, as a delegator, cannot predict who will make the purchase and therefore cannot identify the delegatee until receiving purchase request from a specific buyer. A passive PRE is required in this scenario.

\section{Definition and Security Model}

\subsection{The definition of passive proxy re-encryption}
In current literatures, (active) PRE assumes that the delegator is honest and active to transfer the decryption right to the specified delegatee, and knows his/her public key before generating re-encryption key for him/her.
However, in digital goods transactions, the delegator is passively receiving request from the delegatee, and may often be motivated to not play fair.
Formally, we define passive PRE by adding two new algorithms $Request$ and $VerifyReKey$, based on the typical PRE schemes $Setup,KeyGen,Encrypt,ReKeyGen,ReEncrypt,Decrypt$.
$Request$ is used by the delegatee to make a request and send his/her public key to the delegator. To prevent forgery from malicious delegators, $VerifyReKey$ is used to verify the re-encryption key.

The passive PRE consists of the following eight algorithms.

\noindent $Setup(l_q)$: This algorithm takes a security parameter $l_q$ as input and outputs the global parameters $param$.

\noindent $KeyGen(param)$: The key generation algorithm generates a public/private key pair ${(pk_i,sk_i)}$ for the delegator $i$, and ${(pk_j,sk_j)}$ for the delegatee $j$.

\noindent $Encrypt(sk_i,m)$: The encryption algorithm takes a private key ${sk_i}$ of the delegator $i$ and a plaintext ${m \in \mathcal{M}}$ as input, and outputs the original ciphertext ${CT_i}$. Here $\mathcal{M}$ denotes the message space.

\noindent $Request(sk_j,pk_i)$: With inputs as ${sk_j}$ of the delegatee $j$ and ${pk_i}$ of the delegator $i$, this algorithm outputs a request $R$ for requesting ciphertext conversion from the delegator $i$ to the delegatee $j$.

\noindent $ReKeyGen(sk_i,R)$: For a request $R$ from the delegatee $j$, the delegator $i$ uses this algorithm to generate re-encryption key $rk_{ij}$.

\noindent $VerifyReKey(rk_{ij},CT_i,R)$: Use this algorithm to verify whether the delegator has honestly calculated the re-encryption key $rk_{ij}$ which can transform $CT_i$ into $CT_j$. This algorithm outputs 1 or 0, which means that $rk_{ij}$ is valid or invalid, respectively.

\noindent $ReEncrypt(CT_i,rk_{ij})$: With the input as the original ciphertext ${CT_i}$ and a re-encryption key $rk_{ij}$, this algorithm transforms $CT_i$ into $CT_j$.

\noindent $Decrypt(sk_i/sk_j,CT_i/CT_j)$: The decryption algorithm takes a private key $sk_i$ or $sk_j$ and an original or transformed ciphertext $CT_i$ or $CT_j$ as input, and outputs a message ${m \in \mathcal{M}}$ or the error symbol $\bot$.

In order to simplify the notation, we omit the public parameter $param$ as the input of the algorithms. Correctness requires that for any $param$ and ${m \in \mathcal{M}}$, the following probabilities are equal to 1:
\noindent${Pr\!\left[\!Decrypt(sk_i,CT_i)=m\vline \begin{array}{c}
                                                    (sk_i,pk_i)\leftarrow KeyGen(), \\
                                                    CT_i \leftarrow Encrypt(sk_i,m)
                                                  \end{array}\right]}$\\

\noindent${Pr\!\left[\!Decrypt(sk_j,CT_j)=m\vline\begin{array}{c}
                        (sk_i,pk_i)\leftarrow KeyGen(),\\
                        (sk_j,pk_j)\leftarrow KeyGen(),\\
                        CT_i\leftarrow Encrypt(sk_i,m),\\
                        R\leftarrow Request(sk_j,pk_i),\\
                        rk_{ij}\leftarrow ReKeyGen(sk_i,R),\\
                        CT_j\!\leftarrow\!ReEncrypt(CT_i,rk_{ij})
                      \end{array}\!\right]}$\\

Compared to the definition of active PRE, we add $Request$ and $VerifyReKey$ algorithms for the proposed passive PRE.
The $Request$ algorithm is used to make a request by delegatee, and the $VerifyReKey$ algorithm is used to verify the re-encryption key and avoid the delegator forging.

\subsection{Security model}
The game-based definitions for unidirectional PRE schemes are adaptions of the original ciphertext security and the transformed ciphertext security in \cite{ChowWYD10}\cite{LibertV08}.

\noindent \textbf{Definition 1 (Game Template of Chosen-Ciphertext Security).}\\
\noindent \textbf{Setup.} The challenger $\mathcal{C}$ takes a security parameter $l_q$ and runs the GlobalSetup() algorithm to get the system parameters $param$. Then $\mathcal{C}$ runs the KeyGen() algorithm $n_u$ times resulting a list of public/private keys $\mathcal{PK}_{good}$, $\mathcal{SK}_{good}$, and runs the KeyGen() algorithm for $n_c$ times to get a list of corrupted public/private keys $\mathcal{PK}_{corr}$, $\mathcal{SK}_{corr}$. The challenger $\mathcal{C}$ gives the $param$, $\left(\mathcal{PK}_{good}\bigcup \mathcal{PK}_{corr}=\{pk_i\}_{i\in [1,n_u+n_c]}\right)$ and $\mathcal{SK}_{corr}$ to an adversary $\mathcal{A}$. \\
\noindent \textbf{Game Phase 1.} $\mathcal{A}$ adaptively queries to oracles \textbf{OReK, OReE }and \textbf{ODec}.

\begin{itemize}
   \item \textbf{OReq} oracle takes $\left<pk_j,pk_i\right>$ and returns the request $R$.
   \item \textbf{OReK} oracle takes $\left<pk_i,pk_j\right>$, a request $R$, and returns a re-encryption key $rk_{ij}$ .
   \item \textbf{OReE} oracle takes $\left<pk_i,pk_j\right>$, a ciphertext $CT_i$, a re-encryption key $rk_{ij}$, a request $R$, and returns a transformed ciphertext $CT_j$.
   \item \textbf{ODec} oracle takes a public key $pk$ and a ciphertext $CT$ returns the decryption of $CT$ using the private key of $pk$.
\end{itemize}

\noindent \textbf{Challenge.} When $\mathcal{A}$ decides that \textbf{Game Phase 1} is over, it also decides whether it wants to be challenged with a original ciphertext or a transformed ciphertext. $\mathcal{A}$ outputs two equal length plaintexts $m_0$, $m_1\in \mathcal{M}$, and a target public key $pk_{i*}$. Challenger $\mathcal{C}$ flips a random coin $\theta \in \{0,1\}$, and sends a challenge ciphertext $CT^*$ to $\mathcal{A}$ according to $pk_{i*}$ and $m_\theta$.

\noindent \textbf{Game Phase 2.} $\mathcal{A}$ issues queries as in \textbf{Game Phase 1}.\\
\noindent \textbf{Guess.} Finally, $\mathcal{A}$ outputs a guess $\theta^{\prime}\in \{0,1\}$.

The public key specified by $\mathcal{A}$ must be subject to the following restrictions:\\
\noindent 1. The public keys involved in all queries must come from $\mathcal{PK}$.\\
\noindent 2. The target public key $pk_{i*}$ is from $\mathcal{PK}_{good}$, i.e., uncorrupted.

The actual structure of $CT^*$ and the query constraints performed by $\mathcal{A}$ will be defined according to different security notions.

\noindent \textbf{Definition 2 (Original Ciphertext Security).}  The adversary $\mathcal{A}$ plays CCA game with the Challenger $\mathcal{C}$ according to the rules in definition 1. Let the challenge ciphertext of public key $pk_{i*}$ be $CT_{i*} = Encrypt(ski, m_{\theta})$ . In addition, $\mathcal{A}$ is subject to the following additional constraints:

1. OReK($pk_{i*},pk_j$) is only allowed if $pk_j$ came from $\mathcal{PK}_{good}$.

2. If $\mathcal{A}$ want to issue a OReK($pk_i,pk_j,CT_i$) query where $pk_j$ came from $\mathcal{PK}_{corr}$, $(pk_i,CT_i)$ cannot be a derivative of $(pk_{i*},CT_{i*})$. Please note that derivative notion is to be defined later.

3. ODec($pk,CT$) is only allowed if ($pk,CT$) is not a derivative of $(pk_{i*},CT_{i*})$.

\noindent \textbf{Definition 3 (Derivative for Chosen-Ciphertext Security).} Derivative of $(pk_{i*},CT_{i*})$ in the CCA setting is inductively defined in\cite{ShaoC09} as below, which is adopted from the RCCA based definition in\cite{CanettiH07}:

1. Reflexivity: $(pk_{i*},CT_{i*})$ is a  derivative of itself.

2. Derivation by re-encryption: If $\mathcal{A}$ has issued a re-encryption query $\left<pk_i,pk_j,CT_i\right>$ and obtained the resulting re-encryption ciphertext $CT_j$, then $(pk_j,CT_j)$ is a derivative of $(pk_i,CT_i)$.

3. Derivation by re-encryption key: If $\mathcal{A}$ has issued a re-encryption generation query $(pk_i,pk_j)$ to obtain the re-encryption key $rk_{ij}$, and $CT_j=ReEncrypt(CT_i,\widetilde{rk}_{ij})$, then $(pk_j,CT_j)$ is a derivative of $(pk_i,CT_i)$.

\noindent \textbf{Definition 4 (Transformed Ciphertext Security).} For transformed ciphertext, the adversary $\mathcal{A}$ plays the CCA game with the challenger $\mathcal{C}$ as in Definition 1, where $\mathcal{A}$ can also specify the delegator $pk_i$. Then the challenge ciphertext $CT_{j*}$ is created. Specifically, $CT_{j*}=ReEncrypt(Encrypt(sk_i,m_{\theta}),\widetilde{rk}_{i j*})$. In transformed Ciphertext Security, the  only constraints of $\mathcal{A}$ are:

1. ODec($pk_{j*},CT_{j*}$) is not allowed.

2. If $pk_i$ came from $\mathcal{PK}_{corr}$, $\mathcal{C}$ would not return $\widetilde{rk}_{i j*}$ to $\mathcal{A}$ in \textbf{Game Phase 2}.

3. If $\mathcal{A}$ obtained $\widetilde{rk}_{i j*}$, $\mathcal{A}$ cannot choose $pk_i$ as the delegator in the challenge phase.

\noindent \textbf{Definition 5 (CCA Security of a PRE).} The advantage of $\mathcal{A}$ in attacking the PRE scheme is define as $Adv_{PRE, \mathcal{A}}^{IND-PRE-CCA}=\left|Pr\left[\theta^{\prime}=\theta \right]-1/2 \right|$, where the probability is taken over the random coins consumed by the challenger and the adversary. A single-hop unidirectional PRE scheme is defined to be $\left(t, n_{u}, n_{c}, q_{r k}, q_{r e}, q_{d}, \epsilon\right)$-IND-PRE-CCA secure, if for any t-time -IND-PRE-CCA adversary $\mathcal{A}$, who makes at most $q_{r k}$ re-encryption key generation queries, $q_{r e}$ re-encryption queries and $q_{d}$ decryption queries, we have $Adv_{PRE, \mathcal{A}}^{IND-PRE-CCA} \leq \epsilon $.


\section{Proposed Unidirectional Passive PRE without Pairings}
\subsection{Proposed Construction}
Inspired by PRE schemes of Deng et al. \cite{DengWLC08}, Weng et al. \cite{WengDLC10}, and Chow et al. \cite{ChowWYD10}, we construct a unidirectional passive PRE scheme (UPPRE) without pairings, which consists of the following eight algorithms.\\ \\
\noindent $Setup(l_q)$: Given a security parameter $l_q$, two big primes $p$ and $q$ are chosen such that $q|p-1$, and the bit-length of $q$ is $l_q$. Let $g$ be a generator of group $\mathbb{G}$, which is a subgroup of $\mathbb{Z}^*_p$ with order $q$. Three collision-resistant hash functions can be chosen as follows: ${H_1:\{0,1\}^{l_0}\times\{0,1\}^{l_1}\rightarrow \mathbb{Z}^*_q }$, ${H_2:\mathbb{G}\rightarrow\{0,1\}^{l_0+l_1}}$, and ${H_3:\{0,1\}^*\rightarrow \mathbb{Z}^*_q}$. Here $l_0$ and $l_1$ are also security parameters, and the message space $\mathcal{M}$ is $\{0,1\}^{l_0}$. The global parameters are ${param=(p,q,g,H_1,H_2,H_3,l_0,l_1)}$. By default, the following algorithms all require the parameters ${param}$ as input.

\noindent $KeyGen(param)$: For a delegator $i$, this algorithm picks ${x_{i1}\stackrel{\$}\leftarrow\mathbb{Z}^*_q}$ and ${x_{i2}\stackrel{\$}\leftarrow\mathbb{Z}^*_q}$ randomly, and computes the private key ${sk_i=(x_{i1},x_{i2})}$ and public key ${pk_i=(g^{x_{i1}},g^{x_{i2}})}$. In the same way, the private key ${sk_j=(x_{j1},x_{j2})}$ and the public key ${pk_j=(g^{x_{j1}},g^{x_{j2}})}$ for a delegatee $j$ can be generated.

\noindent $Encrypt(sk_i,m)$: On input ${sk_i}$ and plaintext ${m \in \mathcal{M}}$, this algorithm outputs the original ciphertext ${CT_i}$ under $pk_i$ by following the procedure below.

1. Randomly pick ${u \stackrel{\$}\leftarrow\mathbb{Z}^*_q}$, ${w \stackrel{\$}\leftarrow\{0,1\}^{l_1}}$, and compute ${r=H_1(m,w)}$.

2. Compute ${F=H_2(g^r)\oplus(m||w)}$, ${V=g^{H_3[(x_{i1}+F)\cdot x_{i2}]}}$, ${D=V^u}$, ${E=V^r}$, ${s=u+r\cdot H_3(D,E,F)} (mod q)$.

3. Output ciphertext ${CT_i=(D,E,F,V,s)}$.

\noindent $Request(sk_j,pk_i)$: Delegatee $j$ makes a request to delegator $i$. On input ${sk_j}$ of delegatee $j$ and ${pk_i}$ of delegator $i$, this algorithm outputs a request $R$ for transferring the decryption right of $m$ from delegator $i$ to delegatee $j$  by following the procedure below.

1. Randomly pick ${h \stackrel{\$}\leftarrow\mathbb{Z}^*_q}$ and compute ${g_2=g^h}$.

2. Compute ${\varphi=h\cdot pk_{i1}^{x_{j1}}}$ .

3. Output ${R=(\varphi,g_2,pk_{j1})}$.

\noindent $ReKeyGen(sk_i,R)$: For a correct $R$ from delegatee $j$, the delegator $i$ uses this algorithm to generate re-encryption key 
${rk_{ij}}$, or output an error symbol $\bot.$

1. Compute ${h=\frac{\varphi}{pk_{j1}^{x_{i1}}}}$.

2. Check whether ${g^h=g_2}$ holds. If not, output $\bot.$

3. Compute ${rk_{ij}=\frac{h}{H_3[(x_{i1}+F)\cdot x_{i2}]}}$.

4. Output $(rk_{ij},\varphi)$.

\noindent $VerifyReKey(rk_{ij},V,g_2)$: This algorithm aims to verify if the delegator $i$ has honestly calculated the re-encryption key ${rk_{ij}}$ to transform the decryption right. On input $rk_{ij}$, $V$, and $g_2$, this algorithm outputs 1 or 0, which indicates whether  ${rk_{ij}}$ is valid or not, respectively. Specifically, it follows the procedure below.

1. Check whether $V^{rk_{ij}}=g_2$ holds.

2. If not, output 0. Otherwise, output 1.


\noindent $ReEncrypt(CT_i,(rk_{ij},\varphi),g_2)$: It is required that this algorithm can be executed only when $1\leftarrow VerifyReKey()$. On input the original ciphertext ${CT_i}$, re-encryption key $(rk_{ij},\varphi)$, and $g_2$, this algorithm transforms $CT_i$ into $CT_j$, which is a ciphertext under the public key $pk_j$.  It follows the procedure below.

1. Check whether $V^s=D \cdot E^{H_3(D, E, F)}$ holds. If not, output $\bot$.

2. Compute $E^{\prime}=E^{r k_{i j}}$.

3. Output a transformed ciphertext $CT_j=(E^{\prime},F,\varphi,g_2)$.

\noindent $Decrypt(sk_i/sk_j,CT_i/CT_j)$:

On input a private key $sk_i$ and an original ciphertext $CT_i=(D,E,F,V,s)$, the delegator $i$ can use this algorithm to decrypt $CT_i$, and output the plaintext $m$ or error symbol $\bot$.

1. If $V^s=D \cdot E^{H_3(D, E, F)}$ does not hold, output $\bot$. Else, compute $m||w=F\oplus H_2(E^{\frac{1}{H_{3}\left[\left(x_{i1}+F\right) \cdot x_{i2}\right]}})$.

2. If $E=V^{H_1(m,w)}$ holds, return $m$. Otherwise, output $\bot$.

On input the private key $sk_j$ and a transformed ciphertext $CT_j=(E^{\prime},F,\varphi,g_2)$, the delegatee $j$ can use this algorithm to decrypt $CT_j$, and it outputs the plaintext $m$ or an error symbol $\bot$.

1. Compute $m||w=F\oplus H_2({E^{\prime}}^{\frac{pk_{i1}^{x_{j1}}}{\varphi}})$.

2. If $E^{\prime}=\left(g_{2}\right)^{H_{1}(m, w)}$ holds, return $m$. Otherwise, output $\bot.$

\subsection{Security analysis}
\subsubsection{Complexity Assumptions}
\

\textbf{Definition 6 (Divisible Computation Diffie-Hellman (DCDH) Problem).} Let $\mathbb{G}$ be a cyclic multiplicative group with prime order $q$. The DCDH problem in $\mathbb{G}$ is, given ($g$,$g^a$,$g^b$) $\in{\mathbb{G}}^3$ with $a$, $b{ \stackrel{\$}\leftarrow\mathbb{Z}^*_q}$, to compute $g^\frac{b}{a}$.

\textbf{Definition 7 (DCDH Assumption).} For an algorithm $\mathcal{B}$, its advantage in solving the DCDH problem is defined as $Adv_{\mathcal{B}}^{DCDH} \triangleq Pr\left[\mathcal{B}(g,g^a,g^b)=g^{ab} \right]$, where the probability is taken over the random choices of a, b and those made by $\mathcal{B}$. We say that the $(t,\epsilon)$-DCDH assumption holds in $\mathbb{G}$ if no t-time algorithm $\mathcal{B}$ has advantage at least $\epsilon$ in solving the DCDH problem in $\mathbb{G}$.

\subsubsection{Preliminaries for the Security Proofs}
\

Given an adversary $\mathcal{A}$, who asks at most $q_{H_i}$ random oracle quires to $H_i$ with $i \in$ \{1,2,3\}, and breaks the $(t,n_u,n_c,q_{rk},q_{re},q_d,\epsilon)$-IND-UPPRE-CCA security of our scheme, we will show how to construct a polynomial time algorithm $\mathcal{B}$, which can break the DCDH assumption in $\mathbb{G}$ or the existential unforgeability against chosen message attack (EUF-CMA) of the Schnorr signature with non-negligible advantage. For a cleaner proof, we assume that Schnorr signature is EUF-CMA secure.

The adversary $\mathcal{A}$ who attacks the original ciphertext security is denoted by ${\mathcal{A}}_{orig}$, and the $\mathcal{A}$ who attacks the transformed ciphertext security is denoted by ${\mathcal{A}}_{tran}$. The corresponding algorithms $\mathcal{B}$ are $\mathcal{B}_{orig}$ and $\mathcal{B}_{tran}$ respectively. Our proofs are given in the random oracle model and algorithm $\mathcal{B}$ will simulate the random oracles. $\mathcal{B}$ gives ($q$,$p$,$g$,$H_1$,$H_2$,$H_3$,$l_0$,$l_1$) to $\mathcal{A}$. $H_1$, $H_2$ and $H_3$ are random oracles controlled by $\mathcal{B}$. $\mathcal{B}$ maintains four hash lists ${H_i}^{list}$ with $i \in$ \{1,2,3\}, which are initially empty, and responds the random oracles queries as follow:

\begin{itemize}
   \item $H_1(m,w)$: If this query has appeared on the ${H_1}^{list}$ in a tuple $(m,w,r)$, return the predefined value $r$. Otherwise, randomly pick ${r \stackrel{\$}\leftarrow\mathbb{Z}^*_q}$, add the tuple $(m,w,r)$ to the list ${H_1}^{list}$ and respond with $H_1(m,w)=r$.
   \item $H_2(\rho)$: If this query has appeared on the ${H_2}^{list}$ in a tuple $(\rho,\beta)$, return the predefined value $\beta$. Otherwise, randomly pick $\beta \stackrel{\$}\leftarrow\{0,1\}^{l_0+l_1}$, add the tuple $(\rho,\beta)$ to the list ${H_2}^{list}$ and respond with $H_2(\rho)=\beta$.
   \item $H_3(D,E,F)$: If this query has appeared on the ${H_3}^{list}$ in a tuple $(D,E,F,\gamma)$, return the predefined value $\gamma$. Otherwise, randomly pick ${\gamma \stackrel{\$}\leftarrow\mathbb{Z}^*_q}$, add the tuple $(D,E,F,\gamma)$ to the list ${H_3}^{list}$ and respond with $H_3(D,E,F)=\gamma$. If there is only one input value $X$ when querying, the response is the same as above. For example $H_3(X)$, if this query has appeared on the ${H_3}^{list}$ in a tuple $(X,\gamma)$, return the predefined value $\gamma$. Otherwise, randomly pick ${\gamma \stackrel{\$}\leftarrow\mathbb{Z}^*_q}$, add the tuple $(X,\gamma)$ to the list ${H_3}^{list}$ and respond with $H_3(X)=\gamma$.
\end{itemize}

$\mathcal{B}$ maintains three lists $K^{list}$, $Req^{list}$, and $ReK^{list}$ which stores the list of public/private key pairs, request $R$, re-encryption key generated respectively.

\subsubsection{Original Ciphertext Security}
\

\noindent \textbf{Lemma 1.} The proposed scheme is CCA-secure for the original ciphertext under the DCDH assumption. If a $(t,\varepsilon)$ IND-UPPRE-CCA adversary $\mathcal{A}_{orig}$ with an advantage $\varepsilon$ breaks the IND-UPPRE-CCA security of the given scheme in time t, $\mathcal{C}$ can solve the DCDH problem with advantage $\varepsilon'$ within time $t'$ where
\begin{align*}
\varepsilon'&\geq \frac{1}{q_{H_2}} \left(\frac{\varepsilon}{e(1+q_{rk})}-\frac{q_{H_1}2^{l_0}+q_{H_3}+(q_{H_1}+q_{H_2})q_d}{2^{l_0+l_1}} -\frac{2q_{re}+2q_d}{q} \right) \\
t'&\leq t+(q_{H_1}+q_{H_2}+q_{H_3}+n_u+n_c+q_{rk}+q_{re}+q_d )O(1)\\
  &\quad+(2n_u+2n_c+2q_{rk}+5q_{re}+2q_d+q_{H_1}q_{re}+(2q_{H_1}\\
  &\quad+2q_{H_2})q_d)t_{exp}
\end{align*}

Note that $e$ is the base of natural logarithm and $t_{exp}$ denotes the time to exponentiate in group $\mathbb{G}$. Here $\mathcal{A}_{orig}$ is subject to the previously described restrictions.

Proof of Lemma 1:

\noindent \textbf{Key generations.} $\mathcal{B}_{orig}$ generates the uncorrupted-keys and corrupted-keys as follows.

Uncorrupted-key generation: $\mathcal{B}_{orig}$ picks ${x_{i1}\stackrel{\$}\leftarrow\mathbb{Z}^*_q}$ and ${x_{i2}\stackrel{\$}\leftarrow\mathbb{Z}^*_q}$, and uses Coron's\cite{Coron00} technique to flips a biased coin $c_i \in \{0,1\}$ that yields 1 with probability $\theta$ and 0 otherwise.

\begin{enumerate}[label=(\arabic*)]
   \item If $c_i=1$, it defines ${pk_i=(pk_{i1},pk_{i2})=(g^{x_{i1}},g^{x_{i2}})}$.
   \item If $c_i=0$, it defines ${pk_i=(pk_{i1},pk_{i2})=({(g^{\frac{1}{a}})}^{x_{i1}},{(g^{\frac{1}{a}})}^{x_{i2}})}$.
   \item $\mathcal{B}_{orig}$ add the tuple $(pk_i,x_{i1},x_{i2},c_i)$ to $K^{list}$ and returns $pk_i$ to ${\mathcal{A}}_{orig}$.
\end{enumerate}

Corrupted-key generation: $\mathcal{B}_{orig}$ picks ${x_{j1}\stackrel{\$}\leftarrow\mathbb{Z}^*_q}$ and ${x_{j2}\stackrel{\$}\leftarrow\mathbb{Z}^*_q}$, and defines ${pk_j=(pk_{j1},pk_{j2})=(g^{x_{j1}},g^{x_{j2}})}$, $c_j=-$. Then $\mathcal{B}_{orig}$ add the tuple $(pk_j,x_{j1},x_{j2},c_j)$ to $K^{list}$ and returns $(pk_j,(x_{j1},x_{j2}))$ to ${\mathcal{A}}_{orig}$.

\noindent \textbf{Game Phase 1.} ${\mathcal{A}}_{orig}$ issues a series of queries which $\mathcal{B}_{orig}$ answers ${\mathcal{A}}_{orig}$ as follows:

OReq$(pk_j,pk_i)$:If $Req^{list}$ has an entry for $(pk_j,pk_i)$, return the predefined request $R$ to ${\mathcal{A}}_{orig}$. Otherwise, $\mathcal{B}_{orig}$ acts as follows:

\begin{enumerate}[label=(\arabic*)]
   \item Recover the tuples $(pk_i,x_{i1},x_{i2},c_i)$ and $(pk_j,x_{j1},x_{j2},c_j)$ from $K^{list}$.
   \item If $(c_i=-,c_j=0)$ or $(c_i=1,c_j=0)$, randomly pick ${h\stackrel{\$}\leftarrow\mathbb{Z}^*_q}$ and compute $g_2=g^h$, $\varphi=h\cdot pk_{j1}^{x_{i1}}$.
   \item If $(c_i=0,c_j=0)$, randomly pick ${\varphi\stackrel{\$}\leftarrow\mathbb{Z}^*_q}$, ${g_2\stackrel{\$}\leftarrow\mathbb{Z}^*_q}$, ${h\stackrel{\$}\leftarrow\mathbb{Z}^*_q}$.
   \item Otherwise, ${h\stackrel{\$}\leftarrow\mathbb{Z}^*_q}$ and compute $g_2=g^h$, $\varphi=h\cdot pk_{i1}^{x_{j1}}$.
   \item Let request $R=(\varphi,g_2,pk_j)$, and add tuple $(pk_j,pk_i,R=(\varphi,g_2,pk_j),h)$ to $Req^{list}$. Return $R=(\varphi,g_2,pk_j)$ to ${\mathcal{A}}_{orig}$.
\end{enumerate}

OReK$(pk_j,pk_i,R)$: If $ReK^{list}$ has an entry for $(pk_j,pk_i,R)$,  return the predefined re-encryption key to ${\mathcal{A}}_{orig}$. Otherwise, $\mathcal{B}_{orig}$ acts as follows:

\begin{enumerate}[label=(\arabic*)]
   \item Recover the tuples $(pk_i,x_{i1},x_{i2},c_i)$ and $(pk_j,x_{j1},x_{j2},c_j)$ from $K^{list}$.
   \item If $R=(\varphi,g_2,pk_j)$ does not exist in $Req^{list}$, $\mathcal{B}_{orig}$ checks the $R$ according to the following cases:
     \begin{itemize}
      \item $(c_i\neq 0)$: Compute ${h=\frac{\varphi}{pk_{j1}^{x_{i1}}}}$. Check whether $g_2=g^h$ holds. If not, output $\bot$ and abort. Otherwise, add $(pk_j,pk_i,R=(\varphi,g_2,pk_j),h)$ to $Req^{list}$.
      \item $(c_j\neq 0)$: Compute ${h=\frac{\varphi}{pk_{i1}^{x_{j1}}}}$. Check whether $g_2=g^h$ holds. If not, output $\bot$ and abort. Otherwise, add $(pk_j,pk_i,R=(\varphi,g_2,pk_j),h)$ to $Req^{list}$.
      \item $(c_i=0,c_j=0)$: Output $\bot$ and abort.
     \end{itemize}
   \item If $R=(\varphi,g_2,pk_j)$ exists in $Req^{list}$, $\mathcal{B}_{orig}$ recovers $(pk_j,pk_i,R=(\varphi,g_2,pk_j),h)$ from $Req^{list}$ and constructs $rk_{ij}$ according to the following cases:
     \begin{itemize}
      \item $(c_i=1$or$c_i=-)$: Compute ${rk_{ij}=\frac{h}{H_3[(x_{i1}+F)\cdot x_{i2}]}}$ and let $\tau=1$.
      \item $(c_i=0,c_j=1)$ or$(c_i=0,c_j=0)$: Let ${rk_{ij}\stackrel{\$}\leftarrow\mathbb{Z}^*_q}$ and $\tau=0$.
      \item $(c_i=0,c_j=-)$: Output $\bot$ and abort.
     \end{itemize}
   \item If $\mathcal{B}_{orig}$ does not abort, add tuple $(pk_i,pk_j,(rk_{ij},\varphi),h,R,\tau)$ to $ReK^{list}$, and return $(rk_{ij},\varphi)$ to ${\mathcal{A}}_{orig}$.
\end{enumerate}

OReE$(pk_i,pk_j,CT_i,rk_{ij},R)$: $\mathcal{B}_{orig}$ runs algorithm $VerifyReKey(rk_{ij},V,g_2)$ to check the validity of $rk_{ij}$. If $rk_{ij}$ is invalid, $\mathcal{B}_{orig}$ outputs $\bot$. Otherwise, $\mathcal{B}_{orig}$ acts as follows:

\begin{enumerate}[label=(\arabic*)]
   \item If $V^s\neq D \cdot E^{H_3(D, E, F)}$, output $\bot$ and abort, since $CT_i$ is invalid.
   \item Recover the tuples $(pk_i,x_{i1},x_{i2},c_i)$ and $(pk_j,x_{j1},x_{j2},c_j)$ from $K^{list}$.
   \item $\mathcal{B}_{orig}$ constructs transformed ciphertext $CT_j$ according to the following cases:
     \begin{itemize}
      \item $(c_i\neq0,c_j\neq-)$: Run $ReEncrypt(CT_i,rk_{ij},g_2)$ to generate the transformed ciphertext $CT_j$, and return $CT_j$ to ${\mathcal{A}}_{orig}$.
      \item $(c_i=0,c_j=-)$: Search for the tuple $(m,w,r)\in {H_1}^{list}$ such that $V^r=E$. If there exists no such tuple, return $\bot$(This corresponds to the event $REErr$). If there exists such tuple, run $ReEncrypt(CT_i,rk_{ij},g_2)$ to generate the $CT_j$, and return $CT_j$ to ${\mathcal{A}}_{orig}$.
     \end{itemize}
\end{enumerate}

OReE$(pk_i,CT_i)$:

\begin{enumerate}[label=(\arabic*)]
   \item Recover the tuple $(pk_i,x_{i1},x_{i2},c_i)$ from $K^{list}$. If $c_i=0$ or $c_j=-$, $\mathcal{B}_{orig}$ runs $Decrypt((x_{i1},x_{i2}),c_i)$ to decrypt the original ciphertext(or the transformed ciphertext) $CT_i$, and return the result to ${\mathcal{A}}_{orig}$.
   \item Otherwise, $\mathcal{B}_{orig}$ acts as follows:
   \begin{itemize}
      \item $CT_i$ is an original ciphertext $CT_i=(D,E,F,V,s)$: If $V^s\neq D \cdot E^{H_3(D, E, F)}$, output $\bot$ and abort, since $CT_i$ is invalid. Otherwise, search lists ${H_1}^{list}$ and ${H_2}^{list}$ to see whether there exists $(m,w,r)\in {H_1}^{list}$ and $(\rho,\beta)\in {H_2}^{list}$ such that
          \begin{center}
           $V^r=E$, $\beta \oplus(m||w)=F$ and $\rho=g^r$.
          \end{center}
          If there exists such tuples, return $m$ to ${\mathcal{A}}_{orig}$. Otherwise, return $\bot$.
      \item $CT_i$ is a transformed ciphertext $CT_i=(E',F,\varphi,g_2)$:
        \begin{itemize}
          \item If there exists a tuple $(pk_j,pk_i,(rk_{ij},\varphi),h,R,0)$ in $ReK^{list}$, compute $E=(E')\frac{1}{rk{ij}}$ and search lists ${H_1}^{list}$, ${H_2}^{list}$ and ${H_3}^{list}$ to see whether there exists $(X,\gamma)\in {H_3}^{list}$, $(m,w,r)\in {H_1}^{list}$ and $(\rho,\beta)\in {H_2}^{list}$ such that
            \begin{center}
              $g^{\gamma}=V$, $V^r=E$, $\beta \oplus(m||w)=F$ and $\rho=g^r$.
            \end{center}
                If there exists such tuples, return $m$ to ${\mathcal{A}}_{orig}$. Otherwise, return $\bot$.
          \item If there does not exist a tuple $(pk_j,pk_i,(rk_{ij},\varphi),h,R,0)$ in $ReK^{list}$, search tuple $(pk_i,pk_j,R=(\varphi,g_2,pk_i),h)$ in $Req^{list}$. If among these tuples, there is a $(\varphi,g_2)$ is consistent with the $(\varphi,g_2)$ of this $CT_i$, then extract $h$ from this tuple and search lists ${H_1}^{list}$ and ${H_2}^{list}$ to see whether there exists $(m,w,r)\in {H_1}^{list}$ and $(\rho,\beta)\in {H_2}^{list}$ such that
            \begin{center}
              $g^{r\cdot h}=E'$, $\beta \oplus(m||w)=F$ and $\rho=g^r$.
            \end{center}
                If there exists such tuples, return $m$ to ${\mathcal{A}}_{orig}$. Otherwise, return $\bot$.
        \end{itemize}
     \end{itemize}
\end{enumerate}

\noindent \textbf{Challenge.} When $\mathcal{A}_{orig}$ decides that \textbf{Game Phase 1} is over, it will be challenged with a original ciphertext. $\mathcal{A}_{orig}$ outputs two equal length plaintexts $m_0$, $m_1\in \{0,1\}^{l_0}$ and a target public key $pk_{i*}$. $\mathcal{B}_{orig}$ flips a random coin $\delta \in \{0,1\}$ and recover the tuple $(pk_{i^*},x_{{i1}^*},x_{{i2}^*},c^*)$ from $K^{list}$. $c^*$ must be equal to 1 or 0. According to $pk_{i*}$ and $m_\delta$, $\mathcal{B}_{orig}$ simulates a challenge ciphertext $CT^*$ by the following steps:

\begin{enumerate}[label=(\arabic*)]
  \item If $c^*=1$, $\mathcal{B}_{orig}$ outputs $\bot$ and aborts.
  \item Randomly pick ${e^*,s^*\stackrel{\$}\leftarrow\mathbb{Z}^*_q}$ and compute $V^*=(g^{\frac{1}{a}})^{e^*}$, $D^*=\frac{(g^{\frac{1}{a}})^{e^*\cdot s^*}}{(g^{\frac{1}{b}})^{e^*\cdot e^*}}$, $E^*=(g^{\frac{1}{b}})^{e^*}$.
  \item Randomly pick $F^*\stackrel{\$}\leftarrow\{0,1\}^{l_0+l_1}$ and define $H_3(D^*,E^*,F^*)=e^*$. Randomly pick $w^*\stackrel{\$}\leftarrow\{0,1\}^{l_1}$ and define $H_1(m_\delta,w^*)=\frac{a}{b}$, $H_2(g^{\frac{a}{b}})=(m_\delta,w^*)\oplus F^*$.
  \item Return $CT^*=(D^*,E^*,F^*,V^*,s^*)$ as challenge ciphertext to $\mathcal{A}_{orig}$.
\end{enumerate}

Observe that the challenge ciphertext $CT^*$ is identically distributed as the real original ciphertext from the construction. To illustrate this point, letting $u^*\triangleq s^*-\frac{a}{b}e^*$ and $r^*\triangleq \frac{a}{b}$, we have
\begin{align*}
 D^*&=\frac{(g^{\frac{1}{a}})^{e^*\cdot s^*}}{(g^{\frac{1}{b}})^{e^*\cdot e^*}}=g^{\frac{1}{a}e^*\cdot s^*-\frac{1}{b}e^*\cdot e^*}=(g^{\frac{1}{a}e^*})^{(s^*-\frac{a}{b}e^*)} \\
    &=(V^*)^{u^*},\\
 E^*&=(g^{\frac{1}{b}})^{e^*}=(g^{\frac{1}{a}e^*})^{\frac{a}{b}}=(V^*)^{r^*},\\
 F^*&=H_2(g^{\frac{a}{b}})\oplus (m_\delta,w^*)= H_2(g^{r^*})\oplus (m_\delta,w^*),\\
 s^*&=(s^*-\frac{a}{b}e^*)+\frac{a}{b}e^*=u^*+\frac{a}{b}\cdot H_3(D^*,E^*,F^*)\\
    &=u^*+r^*\cdot H_3(D^*,E^*,F^*).
\end{align*}

\noindent \textbf{Game Phase 2.} $\mathcal{A}_{orig}$ continues to issue queries and $\mathcal{B}_{orig}$ responds to these queries for $\mathcal{A}_{orig}$ as in \textbf{Game Phase 1}.

\noindent \textbf{Guess.} Eventually, $\mathcal{A}_{orig}$ return a guess $\delta'\in \{0,1\}$ to $\mathcal{B}_{orig}$. $\mathcal{B}_{orig}$ picks a tuple $(\rho,\beta) from the list {H_2}^{list}$ and outputs $\rho$ as a solution to the given $(g,g^{\frac{1}{a}},g^{\frac{1}{b}})$ DCDH instance.

Probability Analysis: Let $Ask{H_3}^*$ be the event that $\mathcal{A}_{orig}$ queried $(D^*,E^*,F^*)$ to $H_3$ before the Challenge phase. The simulation of $H_3$ is perfect, as long as $Ask{H_3}^*$ did not occur.  Since $F*$ is randomly chosen from $\{0,1\}^{l_0+l_1}$ in the Challenge phase, we have $Pr\!\left[Ask{H_3}^*\right]\leq \frac{q_{H_3}}{2^{l_0+l_1}}$. Let $Ask{H_1}^*$ be the event that $(m_\delta,w^*)$ has been queried to $H_1$, and $Ask{H_2}^*$ be the event that $g^{\frac{a}{b}}$ has been queried to $H_2$. The simulations of $H_1$ and $H_2$ are also perfect, as long as $Ask{H_1}^*$ and $Ask{H_2}^*$ did not occur where $\delta$ and $w^*$ are chosen by $\mathcal{B}_{orig}$ in the Challenge phase.

Let $Abort$ denote the event of $\mathcal{B}_{orig}$'s aborting during the simulation of the re-encryption key queries or in the Challenge phase and $\neg Abort$ be the event that $Abort$ did not occur. We have $Pr\!\left[Ask{H_3}^*\right]\leq \frac{1}{e(1+q_{rk})}$. Let $REErr$ be the event that $\mathcal{A}_{orig}$ submitted valid original ciphertexts without querying hash function $H_1$. Let $Valid$ be the event that the original ciphertexts is valid. Let $DErr$ be the event that $Valid|(\neg AskH_1\cup \neg AskH_2)$ happens during the entire simulation. Since $\mathcal{A}_{orig}$ issues at most $q_d$ decryption oracles, we have $Pr\!\left[DErr\right]\leq \frac{(q_{H_1}+q_{H_2})q_d}{2^{l_0+l_1}}+\frac{2q_d}{q}$. Let $Err$ be event $(Ask{H_1}^*\cup Ask{H_2}^*\cup Ask{H_3}^*\cup REErr \cup DErr )|Abort$. If $Err$ does not happen, due to the randomness of the output of the random oracle $H_2$, it is clear that $\mathcal{A}_{orig}$ cannot gain any advantage greater than $\frac{1}{2}$ in guessing $\delta$. According to the conclusion of scheme \cite{ChowWYD10}, we have
\begin{align*}
&Pr\![\delta=\delta']\geq \frac{1}{2}-\frac{1}{2}Pr\![Err], \\
&\varepsilon=| 2Pr\![\delta=\delta']-1 | \leq Pr\![Err]=Pr\![(Ask{H_1}^*\cup Ask{H_2}^* \\
&\quad\cup Ask{H_3}^*\cup REErr \cup DErr)|\neg Abort]\\
&\quad\leq (Pr[Ask{H_1}^*]+Pr[Ask{H_2}^*]+Pr[Ask{H_3}^*] \\
&\quad+Pr[REErr]+Pr[DErr])/Pr[\neg Abort]
\end{align*}

Since $\mathcal{B}_{orig}$ picks $w\stackrel{\$}\leftarrow\{0,1\}^{l_1}$ which is hidden by the ``one-time pad'' given by $H_2$, and $Pr[Ask{H_1}^*]\leq \frac{q_{H_1}}{2^{l_1}}$, we have
\begin{align*}
&Pr[Ask{H_2}^*]\geq Pr[\neg Abort]\cdot \varepsilon-Pr[Ask{H_1}^*]-Pr[Ask{H_3}^*]\\
&\quad-Pr[DErr]-Pr[REErr] \\
& \geq \frac{1}{e(1+q_{rk})}-\frac{q_{H_1}2^{l_0}+q_{H_3}+(q_{H_1}+q_{H_2})q_d}{2^{l_0+l_1}} -\frac{2q_{re}+2q_d}{q}.
\end{align*}

If $Ask{H_2}^*$ happens, $\mathcal{B}_{orig}$ will be able to solve DCDH instance. Therefore, we have

\begin{align*}
\varepsilon'&\geq \frac{1}{q_{H_2}}Pr[Ask{H_2}^*] \\
&\geq\frac{1}{q_{H_2}}\left(\frac{\varepsilon}{e(1+q_{rk})}-\frac{q_{H_1}2^{l_0}+q_{H_3}+(q_{H_1}+q_{H_2})q_d}{2^{l_0+l_1}} -\frac{2q_{re}+2q_d}{q} \right)
\end{align*}

From the description of the simulation, the running time of $\mathcal{B}_{orig}$ can be bounded by
\begin{align*}
t'&\leq t+(q_{H_1}+q_{H_2}+q_{H_3}+n_u+n_c+q_{rk}+q_{re}+q_d )O(1)\\
  &\quad+(2n_u+2n_c+2q_{rk}+5q_{re}+2q_d+q_{H_1}q_{re}+(2q_{H_1}\\
  &\quad+2q_{H_2})q_d)t_{exp}
\end{align*}

This completes the proof of Lemma 1.

\subsection{Theoretical analysis}
In theory, we briefly compare the proposed scheme with the most well known PRE schemes, including Weng et al. \cite{WengDLC10}, Shao et al. \cite{ShaoC09}, Chow et al. \cite{ChowWYD10}, Wang et al. \cite{WangC13},  Libert et al. \cite{LibertV08}, and Selvi et al. \cite{SelviPR17}.

Table I shows the performance comparisons of several PRE schemes. First, except for \cite{WengDLC10}, the other schemes are unidirectional, which are more suitable for fair trade transactions. Second, for the first time, a passive PRE is defined and proposed in this paper. Third, there is no pairing computation used in \cite{WengDLC10}, \cite{ShaoC09}, \cite{ChowWYD10}, \cite{SelviPR17}, and the proposed scheme, which is more practical for resource-constrained environment, like smart contracts. In addition, the PRE schemes proposed in \cite{ShaoC09}, \cite{ChowWYD10}, and \cite{SelviPR17} can resist collusion attacks, which can avoid the exposure of the delegator's private key if the proxy colludes with the delegatee. However, all ciphtertext under this private key can be decrypted illegally. The proposed scheme is backward secure under the attacks launched by colluded proxy and delegatee, which means the private key of delegator cannot be leaked, and the other ciphertext under this private key cannot be decrypted. Last but not least, except for \cite{WengDLC10} and \cite{ShaoC09}, the other schemes are CCA-secure based on the related difficulty assumptions,  among which only \cite{WangC13} and \cite{LibertV08} are secure in the standard model.
In summary, the proposed scheme is the only passive and unidirectional PRE scheme, which is collusion-secure and CCA-secure in the random model based on CDH assumption.

\begin{table*}[]
\renewcommand\arraystretch{1.5}
\centering
\caption{The Performance Comparisons of Several PRE Schemes}
\begin{tabularx}{\linewidth}{lllllllll}
\hline
\multicolumn{2}{l}{Schemes}  & \cite{WengDLC10} & \cite{ShaoC09} & \cite{ChowWYD10} &  \cite{WangC13} &  \cite{LibertV08} &  \cite{SelviPR17} & UPPRE   \\ \hline
\multicolumn{2}{l}{Unidirectional/Bidirectional} & Bidirectional  &  Unidirectional  &  Unidirectional    &  Unidirectional  &    Unidirectional    &    Unidirectional       & Unidirectional \\
\multicolumn{2}{l}{Active/Passive}               &    Active      &    Active        &    Active          &    Active        &    Active            &    Active                  &    Passive      \\
\multicolumn{2}{l}{Pairings}                        &      No        &      No          &       No           &        Yes       &       Yes            &         No              &    No          \\
\multicolumn{2}{l}{Collusion resistance}            &      No        &      Yes         &       Yes          &  Uncertain       &       Uncertain      &         Yes             &    Yes and Forword Secure  \\
\multicolumn{2}{l}{Security level}                  &    Not CCA     &    Not CCA       &           CCA      &        CCA       &        RCCA          &        CCA              &    CCA         \\
\multicolumn{2}{l}{Standard model}                  &      No        &      No          &       No           &        Yes       &        Yes           &         No              &    No          \\
\multicolumn{2}{l}{Assumptions}                     &      CDH       &      DDH         &       CDH          &       DBDH       &        3-QDBDH       &        CDH, DCDH        &    CDH         \\ \hline
\end{tabularx}
\label{tbl:table1}
\end{table*}

We choose PRE schemes that are also unidirectional (i.e. \cite{ChowWYD10}\cite{ShaoC09}\cite{WangC13}\cite{LibertV08}\cite{SelviPR17}) to compare with the proposed scheme on efficiency.
We analyze theoretical computational costs of different algorithms in these PRE schemes.
In particular, $t_{exp}$, $t_p$, $t_s$ and $t_v$ denote the computational costs of an exponentiation, a bilinear pairings, a one-time signature, and a verification, respectively.
$Orig.CT$ and $Trans.CT$ denote the original ciphertext and the transformed ciphertext, respectively.
$|\mathbb{G}|$,$|\mathbb{Z}_q|$,$|\mathbb{G}_1|$ and $|\mathbb{G}_T|$ denote the bit-length of an element in groups $\mathbb{G}$,$\mathbb{Z}_q$,$\mathbb{G}_1$ and $\mathbb{G}_T$ respectively.
$N_X$ and $N_Y$ are the safe-prime modulus in scheme of Shao et al.\cite{ShaoC09}.
In scheme of Libert et al.\cite{LibertV08}, $|svk|$ and $|\sigma|$ denote the length of a verification key and a strong unforgeable one-time signature, respectively.
In our calculation, the computational cost of $g^{r_1}\cdot g^{r_2}$ or $(g^{r1}\cdot g)^{r_2}$ will be considered as $2t_{exp}$.
As shown in Table II, the efficiency of the proposed scheme is higher than that of scheme \cite{SelviPR17}, and slightly higher than that of scheme \cite{ChowWYD10}. Please note that we specifically add  $Request$ and $VerifyReKey$ algorithms to propose the passive PRE which is more suitable for digital goods transactions.
Schemes of \cite{WangC13} and \cite{LibertV08} are built under the standard oracle model and require bilinear parings, while the proposed PRE scheme is built under random oracle model and does not require parings.

\begin{table*}[]
\renewcommand\arraystretch{1.5}
\caption{The Efficiency Comparisons of Several PRE Schemes}
\begin{tabularx}{\linewidth}{lllllllll}
\hline
\multicolumn{3}{l}{Schemes}& \cite{ShaoC09} &  \cite{ChowWYD10} & \cite{WangC13} &  \cite{LibertV08} & \cite{SelviPR17} & UPPRE \\ \hline
\multirow{7}{*}{Cost}     & \multicolumn{2}{l}{Encrypt}         & 5$t_{exp}$  & 4$t_{exp}$  & 1$t_p$+5$t_{exp}$& 1$t_p$+4$t_{exp}$+1$t_s$&5$t_{exp}$ & 4$t_{exp}$      \\
                          & \multicolumn{2}{l}{Request}      &$\backslash$ &$\backslash$ &$\backslash$ & $\backslash$ &  $\backslash$    & 2$t_{exp}$ \\
                          & \multicolumn{2}{l}{ReKeyGen}        & 2$t_{exp}$  & 2$t_{exp}$  &1$t_p$+6$t_{exp}$& 1$t_{exp}$&  2$t_{exp}$     & 2$t_{exp}$ \\
                          & \multicolumn{2}{l}{VerifyReKey}   &$\backslash$ &$\backslash$ &$\backslash$ &$\backslash$   & $\backslash$   &  1$t_{exp}$ \\
                          & \multicolumn{2}{l}{ReEncrypt}       & 5$t_{exp}$  & 4$t_{exp}$ &5$t_p$+6$t_{exp}$ &2$t_p$+4$t_{exp}$+1$t_v$ &7$t_{exp}$ &  4$t_{exp}$  \\
                          & \multirow{2}{*}{Decrypt}&Orig.CT  & 6$t_{exp}$  & 5$t_{exp}$ &3$t_p$+3$t_{exp}$ &3$t_p$+2$t_{exp}$+1$t_v$ &9$t_{exp}$ &  4$t_{exp}$          \\
                          &                         &Trans.CT & 4$t_{exp}$  & 4$t_{exp}$  &4$t_p$+4$t_{exp}$ &5$t_p$+2$t_{exp}$+1$t_v$ & 4$t_{exp}$ &3$t_{exp}$           \\
\multirow{2}{*}{Length}   & \multicolumn{2}{l}{Orig.CT}         & $2k\!+\!3|N_X^2|\!+|m|$               &$3|\mathbb{G}|+|\mathbb{Z}_q|$ &$2|\mathbb{G}_1|+1|\mathbb{G}_T|$ &$1|svk|\!+\!2|\mathbb{G}_1|\!+\!1|\mathbb{G}_T|\!+\!1|\sigma|$  &$3|\mathbb{G}|+|\mathbb{Z}_q|$ &$4|\mathbb{G}|+|\mathbb{Z}_q|$\\
                          & \multicolumn{2}{l}{Trans.CT}        &$k_1\!+\!3|N_X^2|\!+\!2|N_Y^2|\!+\!|m|$&$2|\mathbb{G}|+2|\mathbb{Z}_q|$&$5|\mathbb{G}_1|+1|\mathbb{G}_T|$ &$1|svk|\!+\!4|\mathbb{G}_1|\!+\!1|\mathbb{G}_T|\!+\!1|\sigma|$  &$4|\mathbb{G}|$                &$4|\mathbb{G}|$                \\  \hline

\end{tabularx}
\label{tbl:table2}
\end{table*}

\subsection{Experimental analysis}
In order to ensure a fair comparison, we chose schemes \cite{ChowWYD10} and \cite{SelviPR17} to compare with the proposed scheme because they are also unidirectional without paring and meet CCA security in the random oracle model. We implement these schemes through java programming language on a computer that consists of an Intel (R) Core (TM) i7-8750 processor and a RAM with total memory of 16GB. In order to facilitate the experiments, we always set $p=2q+1$ and $l_0=l_1$.

\begin{figure*}

\begin{minipage}[t]{0.5\textwidth}
\centering
\includegraphics[width=2.5in]{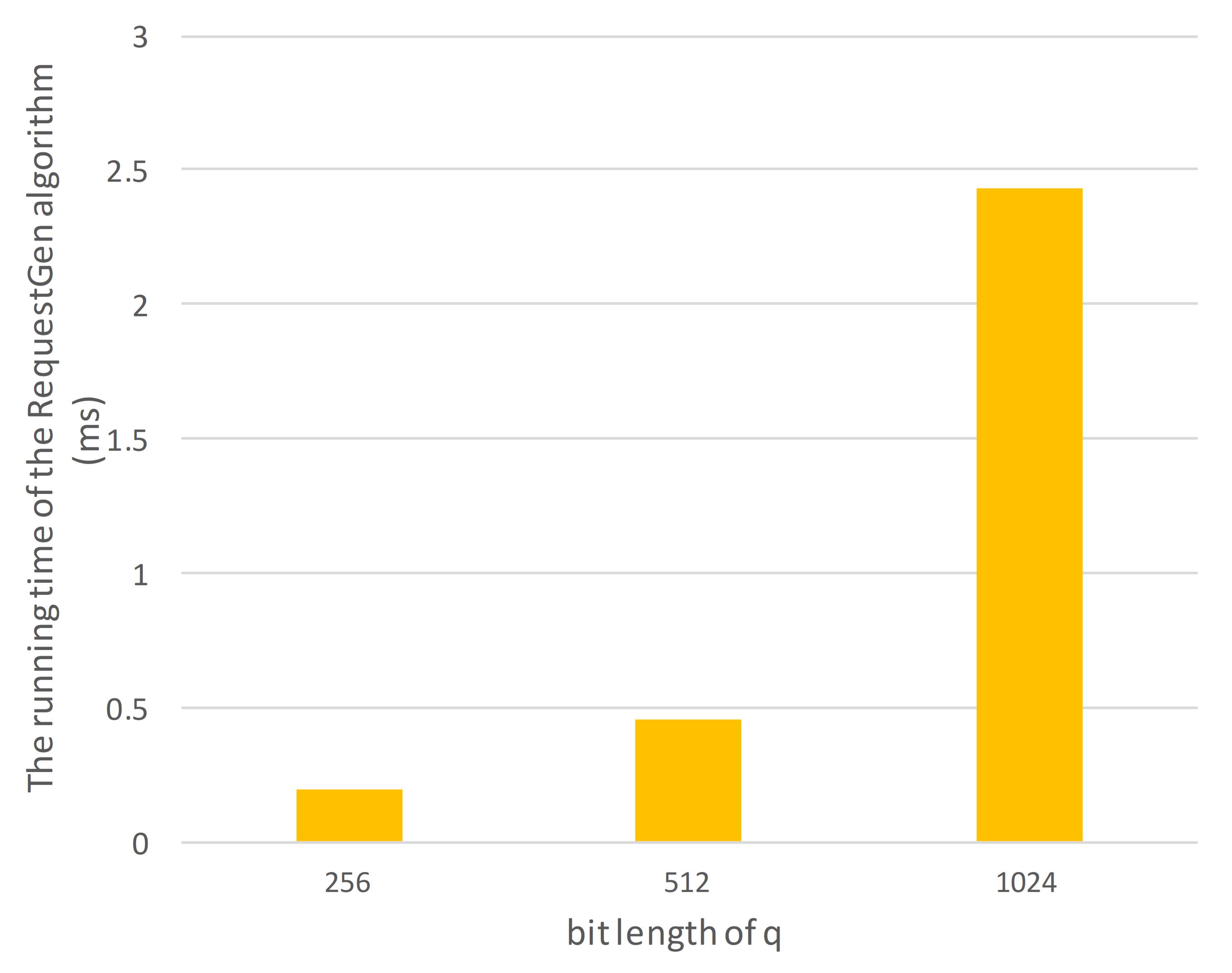}
\caption*{(a) $Request$}
\end{minipage}%
\begin{minipage}[t]{0.5\textwidth}
\centering
\includegraphics[width=2.5in]{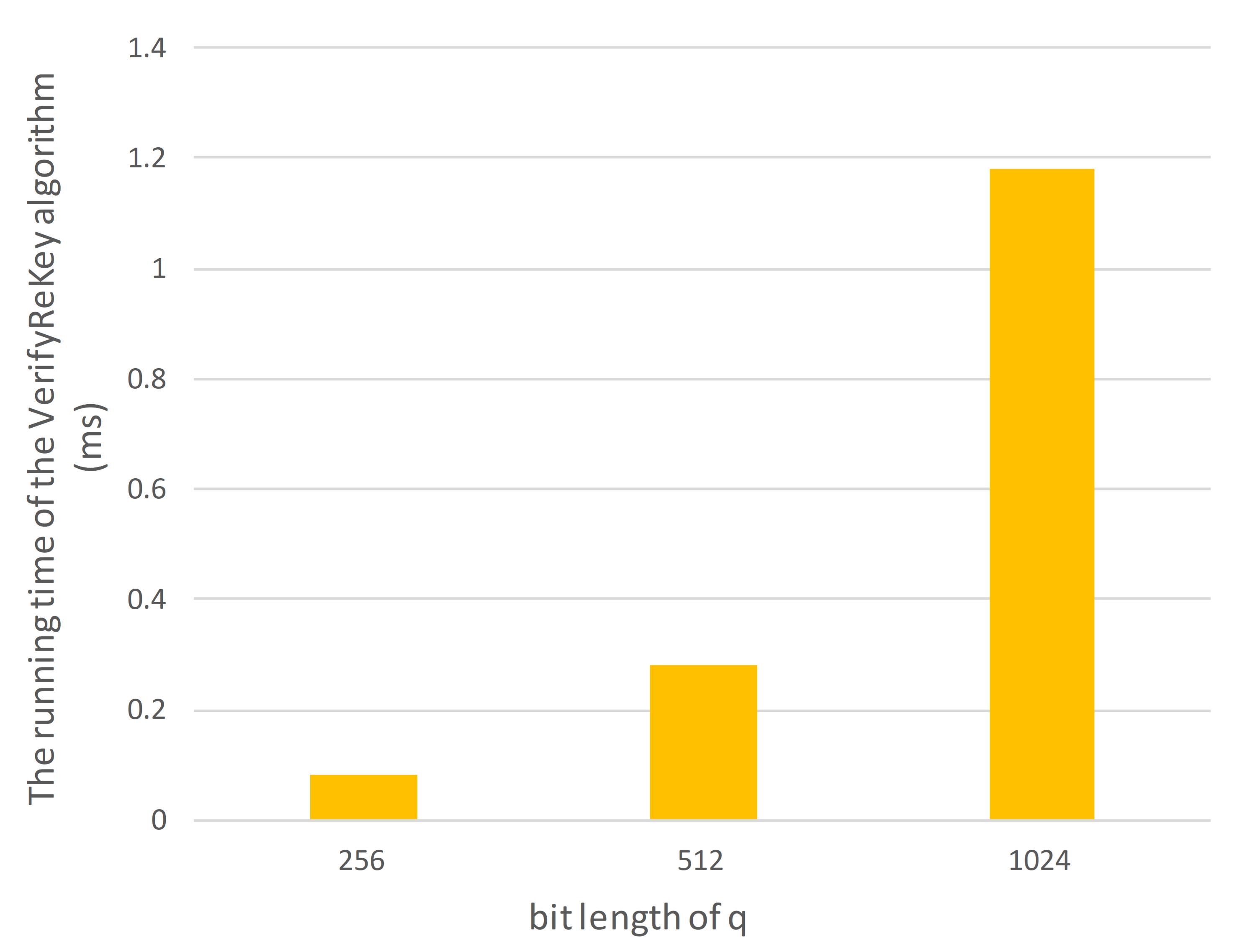}
\caption*{(b) $VerifyReKey$}
\end{minipage}%
\caption{The running time of $Request$ and $VerifyReKey$ algorithms}
\label{fig:RequestGen and VerifyReKey}
\end{figure*}

\begin{figure*}

\begin{minipage}[t]{0.33\textwidth}
\centering
\includegraphics[width=2.2in]{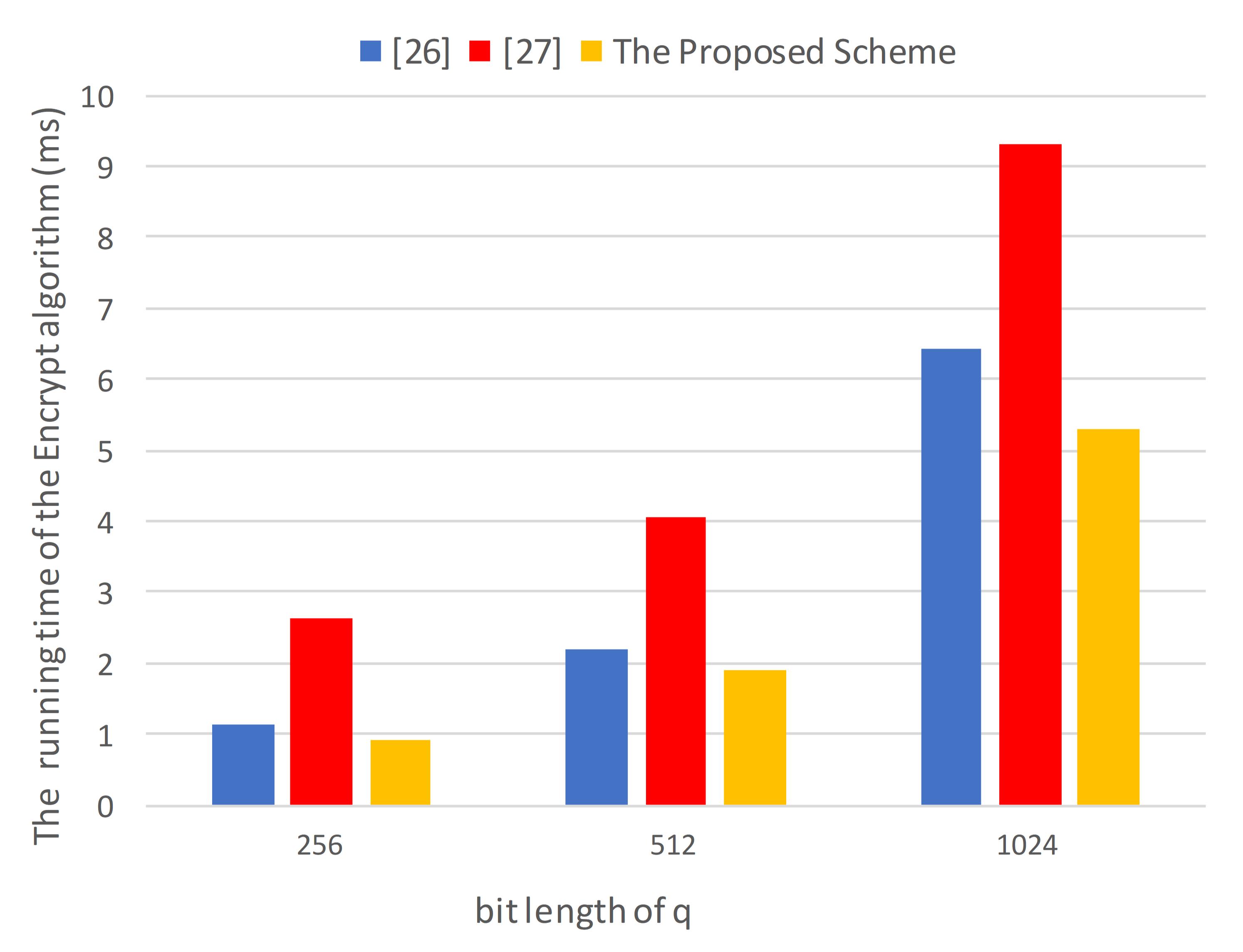}
\caption*{(a) $Encrypt$}
\end{minipage}%
\begin{minipage}[t]{0.33\textwidth}
\centering
\includegraphics[width=2.2in]{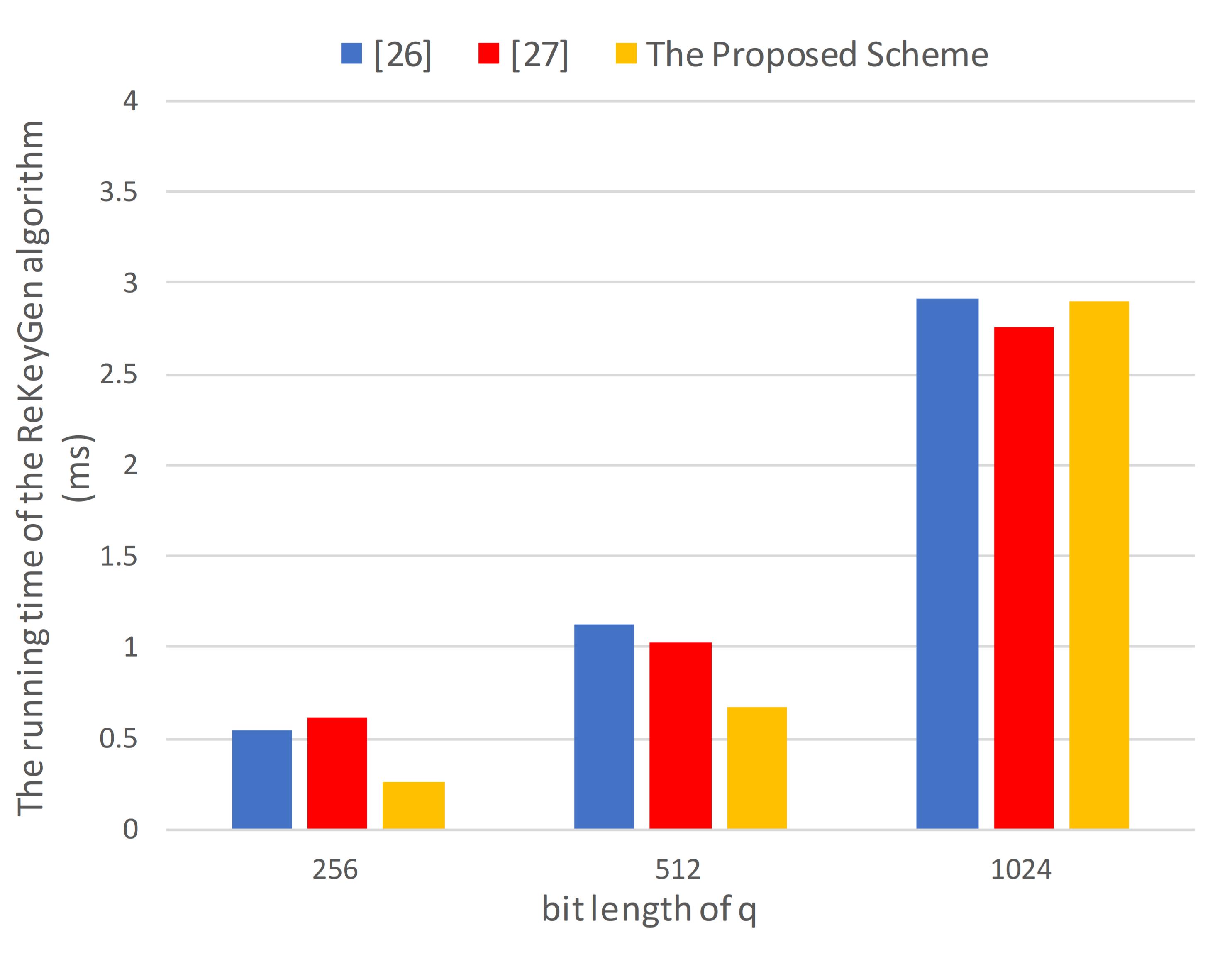}
\caption*{(b) $RekeyGen$}
\end{minipage}%
\begin{minipage}[t]{0.33\textwidth}
\centering
\includegraphics[width=2.2in]{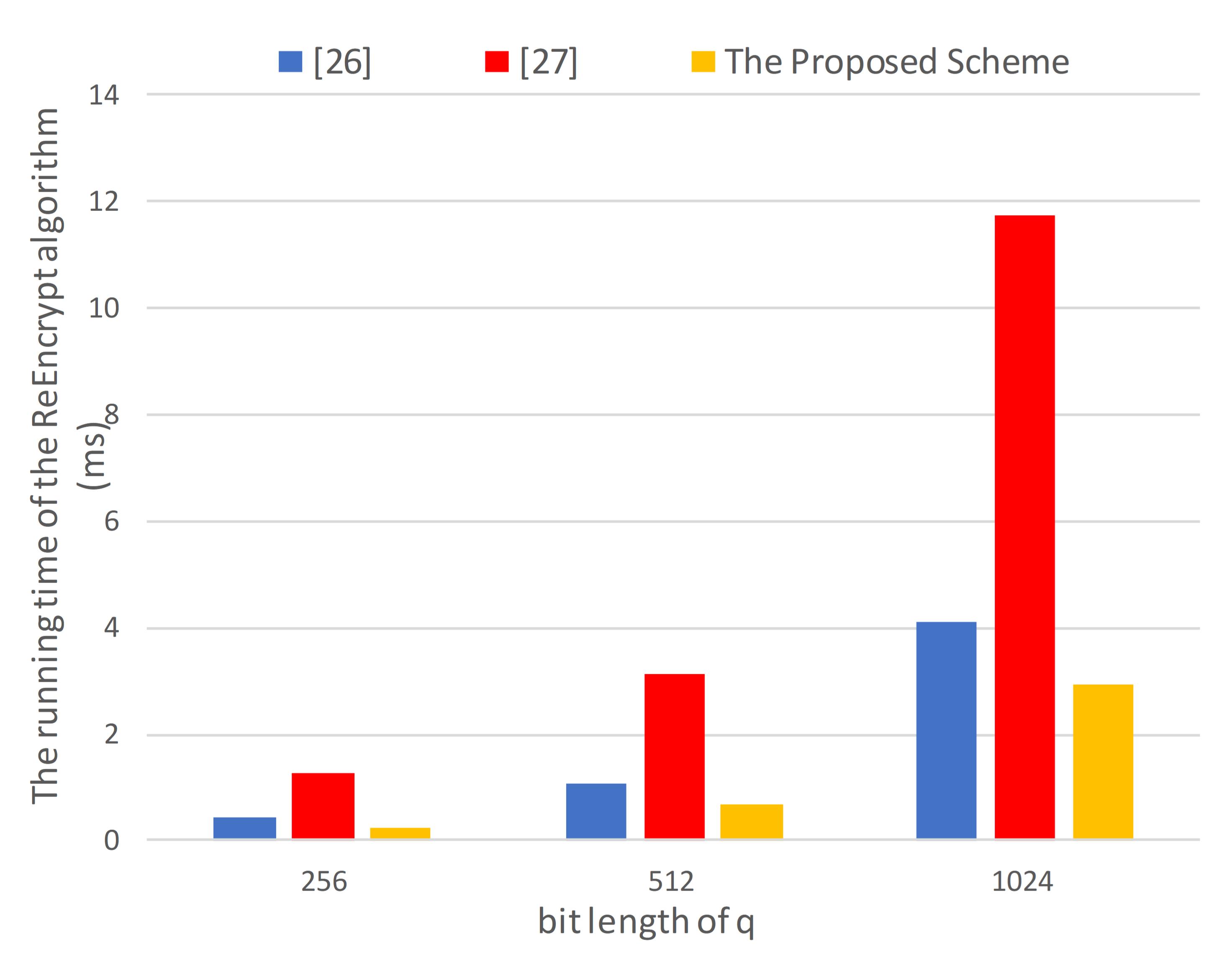}
\caption*{(c) $ReEncrypt$}
\end{minipage}

\vfill

\begin{minipage}[t]{0.33\textwidth}
\centering
\includegraphics[width=2.2in]{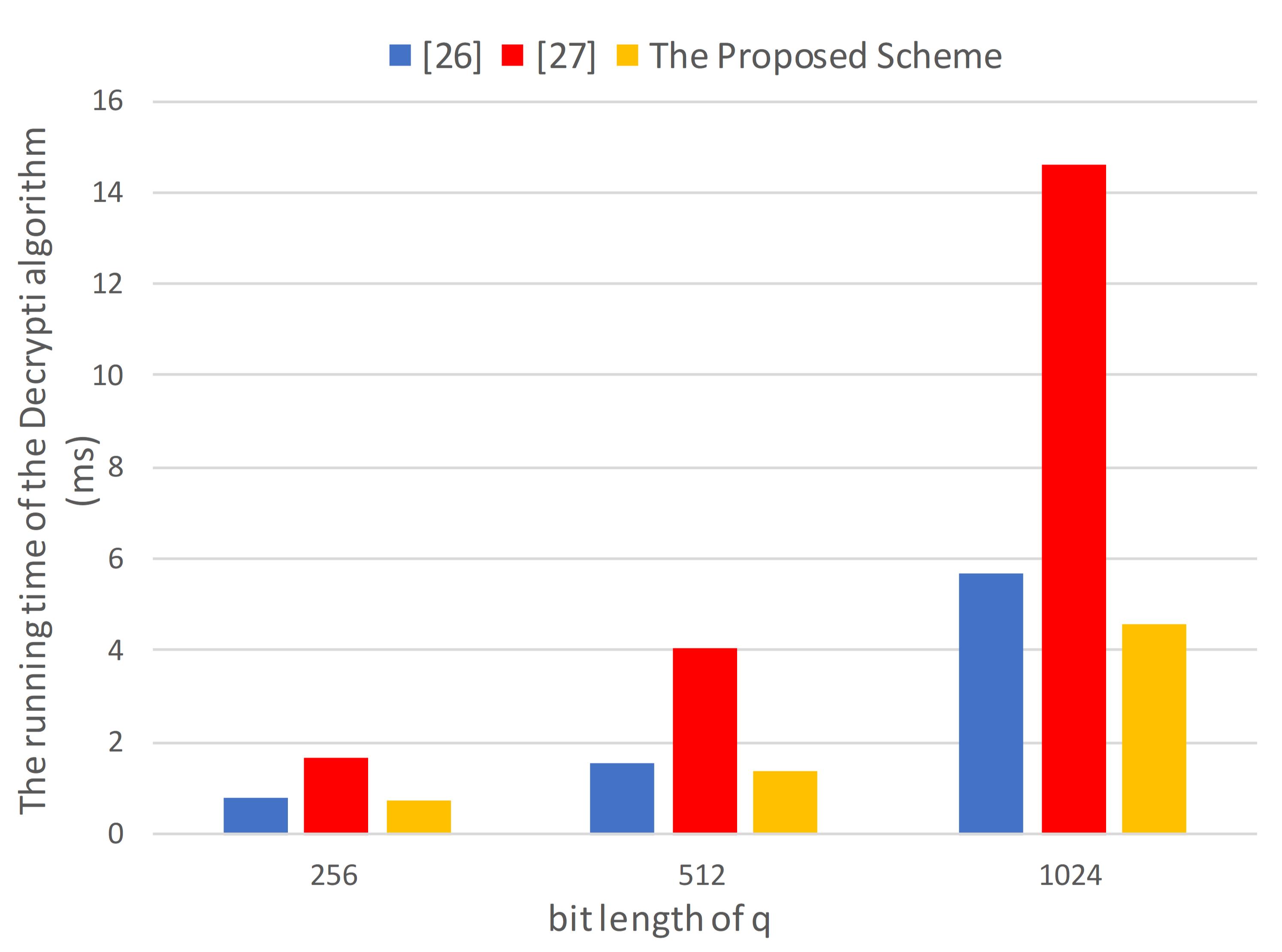}
\caption*{(d) $Decrypt$ for Orig.CT}
\end{minipage}%
\begin{minipage}[t]{0.33\textwidth}
\centering
\includegraphics[width=2.2in]{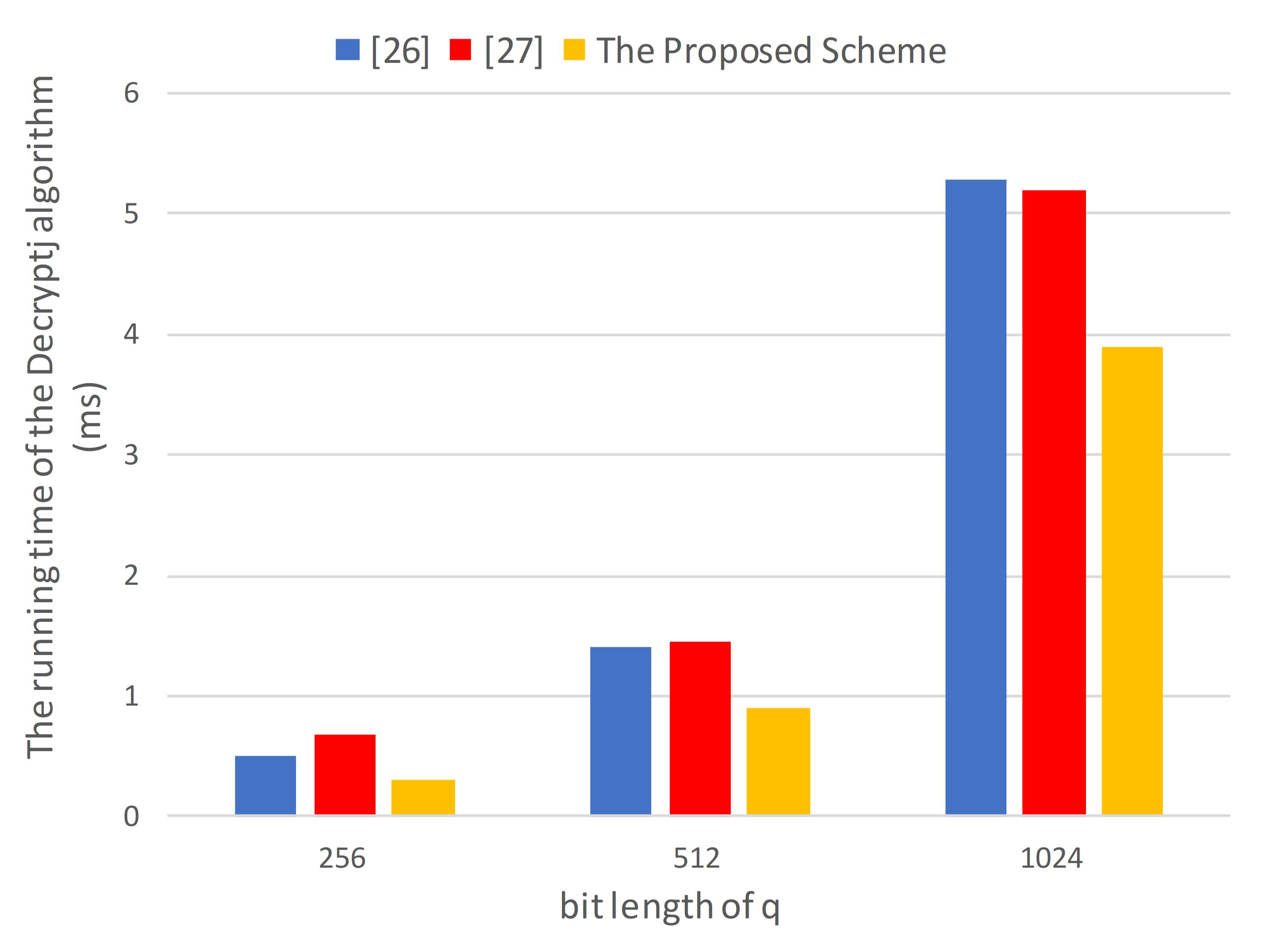}
\caption*{(e) $Decrypt$ for Trans.CT}
\end{minipage}%
\begin{minipage}[t]{0.33\textwidth}
\centering
\includegraphics[width=2.2in]{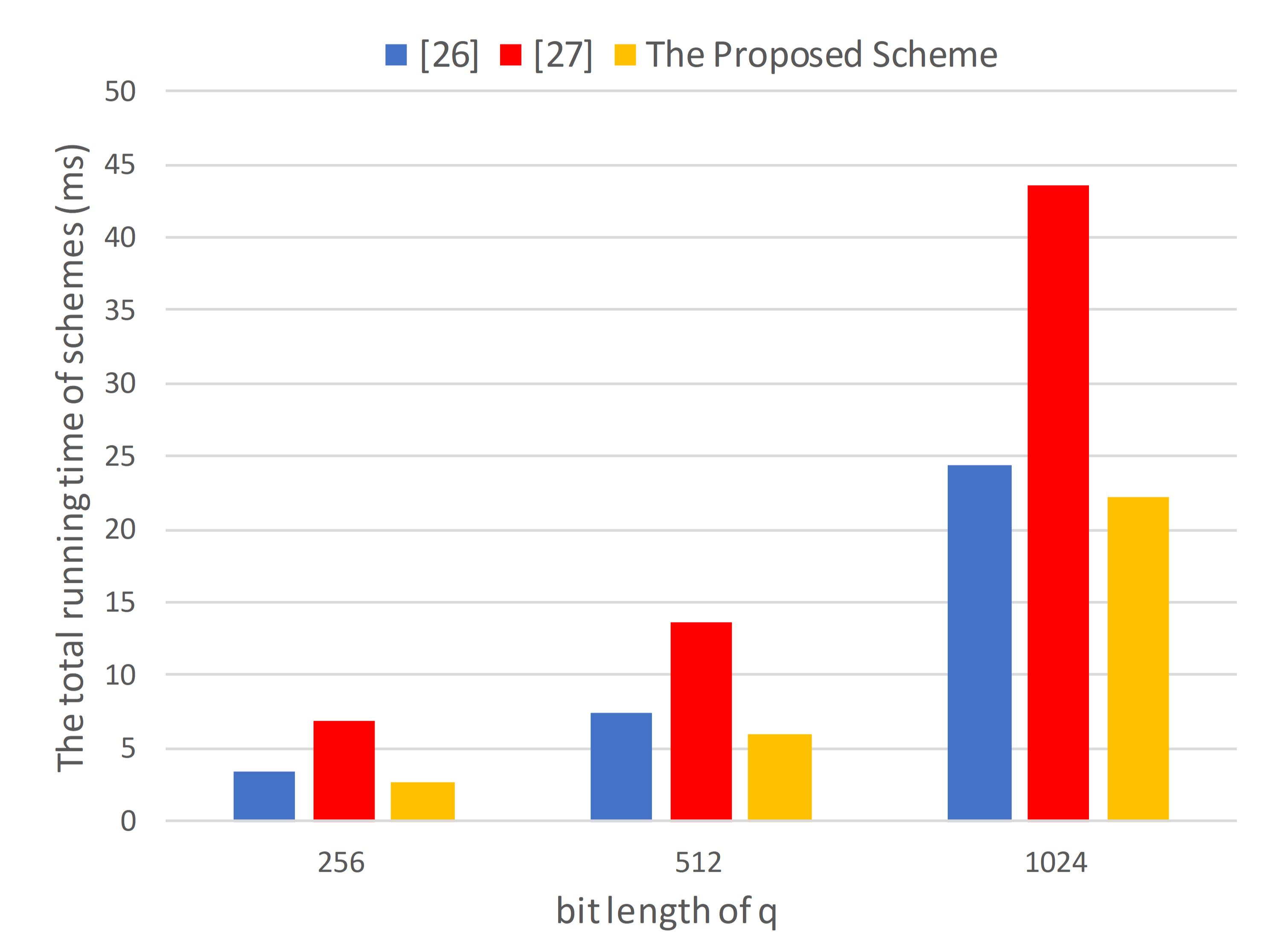}
\caption*{(f) Total}
\end{minipage}
\caption{The running time comparison of schemes}
\label{fig:The running time comparison of schemes}
\end{figure*}

We execute each algorithm 50 times and present the average running time. Let $l_q=256, 512, and 1024$ respectively, and $p=2q+1$, $l_0=l_1=\frac{1}{2}l_q$,
and the experiment results are shown in Fig.\ref{fig:RequestGen and VerifyReKey} and Fig.\ref{fig:The running time comparison of schemes}.
Please note that $Request$ and $VerifyRekey$ algorithms only exist in our scheme.
In general, the proposed scheme has advantages in the running time of each algorithm and has obvious advantages in the total running time.


%
%

\section{The proposed fair trade protocol}

\subsection{Fair trade protocol framework}
\begin{figure}[H]
     \centering
     \includegraphics[scale=0.65]{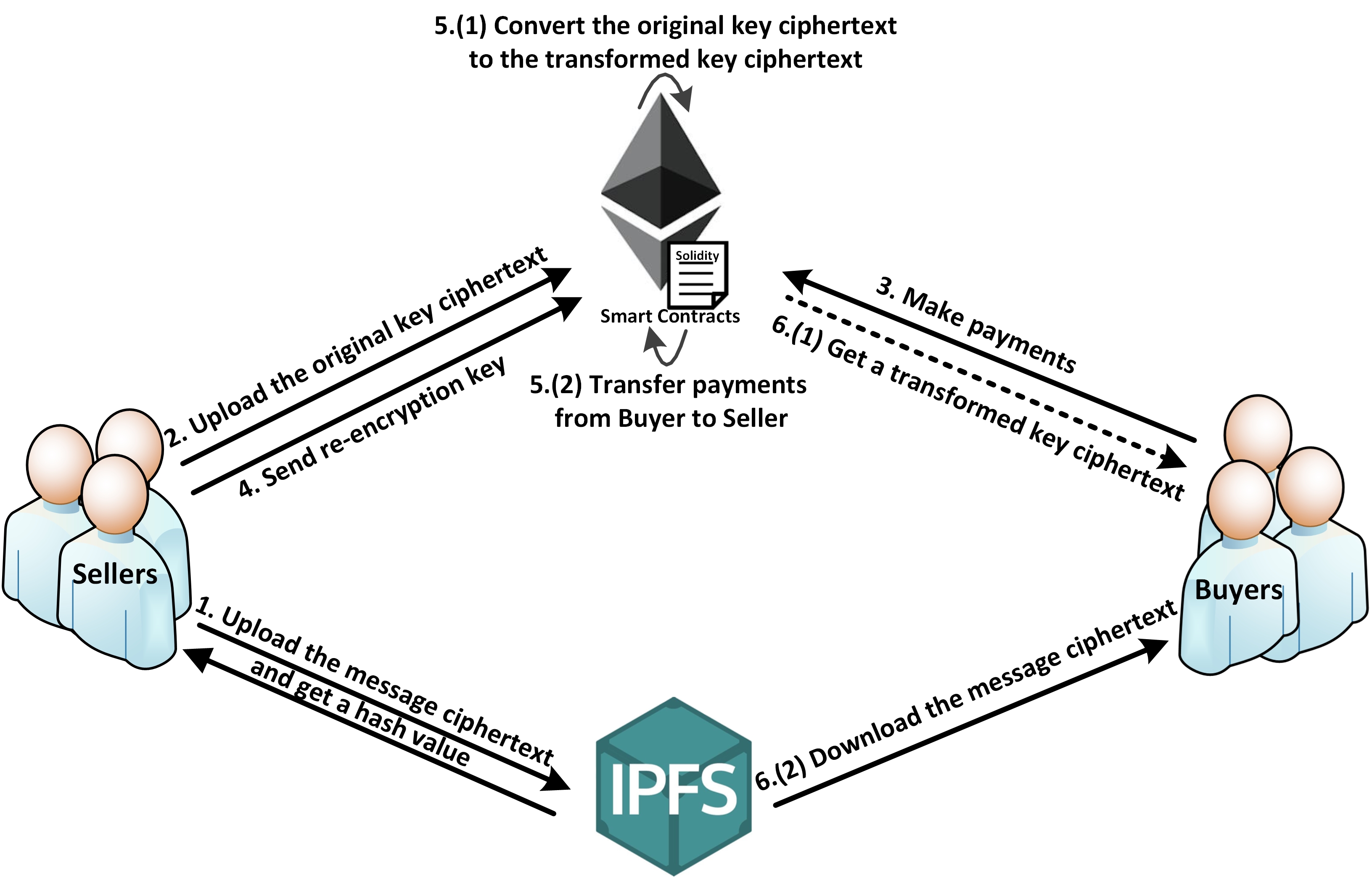}
     \caption{Our fair protocol framework}
     \label{fig:protocol framework}
    \end{figure}

As described above, based on the proposed passive PRE scheme UPPRE, we introduce smart contracts to be the proxy, who is responsible for automatically releasing the decryption right of digital goods after payments. As the data of digital goods may be large, it is difficult to store the data directly on Blockchain.
Therefore, InterPlanetary File System (IPFS), a protocol and peer-to-peer Network for storing and sharing data in a distributed file system, is introduced to store digital goods.
The proposed fair trade protocol has four roles: IPFS, Sellers, Buyers, and Smart contracts. As shown in Fig.\ref{fig:protocol framework}, these parties' responsibilities are described as follows.

\noindent \textbf{IPFS:} Generally, the size of the digital goods is large and it is not suitable to store them directly in smart contracts. We propose to solve this issue by adopting IPFS. In particular, files can be uploaded to IPFS. Then IPFS will return a unique hash value calculated based on the file's content. Anyone who obtains the corresponding hash value can download the file.

\noindent \textbf{Sellers:} A seller is a user who has the right to sell digital goods. For example, the seller can be the producer of digital goods, such as a film company, or the author of an e-book. After uploading the digital goods ciphertext encrypted using symmetric encryption to IPFS, the seller runs the encryption algorithm of UPPRE to encrypt the corresponding symmetric-key, and upload the original key ciphertext to the smart contracts. The seller needs to register a Blockchain (e.g. Ethereum) account to collect payment when the transaction is completed.

\noindent \textbf{Buyers:} A buyer is a user who wants to buy digital goods from a seller. Each buyer also needs to have a Blockchain account to interact with smart contracts and make payments. After the transaction is successfully completed, the buyer will get a transformed key ciphertext converted by smart contracts. Then the buyer can obtain the symmetric key of the corresponding digital goods after decrypting. Finally, the buyer downloads the encrypted digital goods from IPFS, and decrypts it with the retrieved symmetric key.

\noindent \textbf{Smart Contracts:} In this work, we assume that the smart contracts are deployed by the seller. However, the seller cannot cheat by arbitrarily changing the codes or data of the smart contract because they are public and can be verified by any users.
In fact, the codes are executed automatically, without intervention from anyone. As a result, it is easy to ensure that smart contracts are neutral and don't incline to one party in the transaction.

The buyer and seller complete transactions by interacting with smart contracts.
 The smart contracts are responsible for verifying the inputs submitted by the buyer and seller, collecting the buyer's payment and the seller's re-encryption key.
 When the smart contracts determine that a transaction is established, it will convert the original key ciphertext into the transformed key ciphertext, and transfer the owner of the payment from the buyer to the seller.
 When disputes arise, the smart contracts can guarantee the fairness of the transaction through the arbitration protocol.

\subsection{The fair trade protocol}

According to the proposed framework, the whole processing flows of the proposed fair trade protocol is shown in Figure 4, which includes the following phases:
   \begin{figure*}[hb]
     \centering
     \includegraphics[scale=0.58]{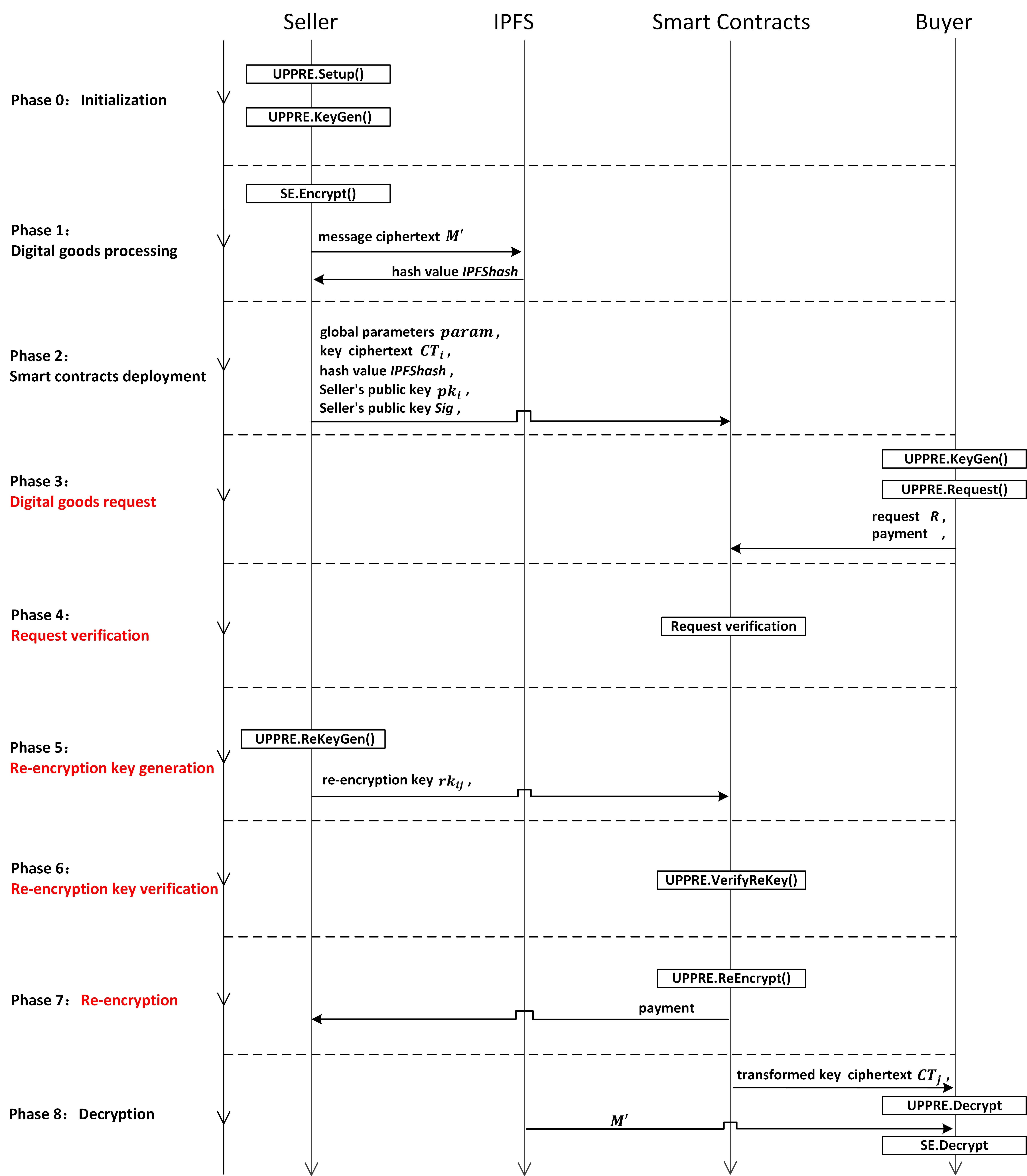}
     \caption{The proposed fair trade protocol process (Phases in red text indicates from beginning to end of a transaction.)}
    \end{figure*}

\textbf{Phase 0: Initialization.}

In this phase, the seller runs the $Setup$ algorithm in UPPRE to generate global parameters, and runs the $KeyGen$ algorithms in UPPRE to generate his/her public/private key pair $(pk_i,sk_i)$.

1) $param \leftarrow UPPRE.Setup(l_q)$.

2) $(pk_i,sk_i)\leftarrow UPPRE.KeyGen(param)$.

\textbf{Phase 1: Digital goods processing.}

Before uploading digital goods $M$ to IPFS, the seller encrypts $M$ by a symmetric encryption algorithm (e.g. AES), denoted as $SE=\{Encrypt,Decrypt\}$, and the used symmetric-key is represented by $K$.

1) $M' \leftarrow SE.Encrypt(M,K)$.

2) Upload the message ciphertext $M'$ to IPFS and get $IPFShash$ of the $M'$, through which others can download $M'$.

\textbf{Phase 2: Smart contracts deployment.}

1) Symmetric-key $K$ is encrypted by the seller using the $Encrypt$ algorithm of UPPRE. $CT_i$ is the ciphertext of $K$. \\
\centerline{$CT_i=(D,E,F,V,s) \leftarrow UPPRE.Encrypt(sk_i,K)$}

2) Generate a signature $Sig$ for $CT_i||IPFShash$ by a digital signature algorithm (e.g. Schnorr Signature).
Note that anyone can verify the validity of the $Sig$, and this signature $Sig$ guarantees that $CT_i||IPFShash$ is consistent with the claimed digital goods.

3) Write $param, pk_i, IPFShash, CT_i, Sig$ into smart contracts, and deploy smart contracts to Ethereum.

\textbf{Phase 3: Digital goods request.}

If a buyer would like to purchase this digital goods, he/she verifies the signature $Sig$ firstly.
If the verification succeeds, the buyer believes that $CT_i||IPFShash$ is consistent with the claimed digital goods, and performs the following steps:

1) Generate his own public/private key pair $(pk_j,sk_j)$:\\
\centerline{$(pk_j,sk_j)\leftarrow UPPRE.KeyGen(param)$}

2) Compute the request $R$:\\
\centerline{$R=(\varphi,g_2,pk_{j})\leftarrow UPPRE.Request(sk_j,pk_i)$}

3) Submit the priced payment and the request $R$ to smart contracts.

\textbf{Phase 4: Request verification.}

The smart contracts maintain a request list $L_R$ to store public keys ever requested.
When receiving a request $R$ from a buyer, smart contracts check whether the $pk_{j}$ has appeared in the list $L_R$.

1) If yes, reject this request and refund the payment, and this transaction is terminated.

2) If no, accept the request $R$ and add $pk_{j}$ to the list $L_R$.

3) Smart contracts trigger an event monitored by the seller, so that the seller will know it is his/her turn.

\textbf{Phase 5: Re-encryption key generation.}

When smart contracts receive a request $R$ from a buyer, the seller generates a re-encryption key $rk_{ij}$ according to the $R$.

1) Run $rk_{ij} \leftarrow UPPRE.ReKeyGen(sk_i,R)$.

2) Send this re-encryption key $rk_{ij}$ to smart contracts.

\textbf{Phase 6: Re-encryption key verification.}

In order to check the validity of $rk_{ij}$, smart contracts use the $VerifyRekey$ algorithm of UPPRE to check.
If smart contracts have not received the re-encryption key at all in a time period $T$,  or if $0 \leftarrow UPPRE.VerifyReKey(rk_{ij},V,g_2)$, the verification fails and the arbitration protocol will be launched. Otherwise, the re-encryption key passes the verification.

\textbf{Phase 7: Re-encryption.}

With the Re-encryption key, smart contracts will transform the ciphertext $CT_i$ to $CT_j$ by the $ReEncrypt$ algorithm of UPPRE.

1) Run $CT_j \leftarrow UPPRE.ReEncrypt(CT_i,rk_{ij},g_2)$.

2) Store $CT_j$ into smart contracts.

3) Transfer the payment to the seller.

Up to this phase, this transaction is completed. The buyer goes to next phase to decrypt the ciphertext of digital goods.

\textbf{Phase 8: Decryption.}

1) The buyer gets the transformed ciphertext $CT_j$ from smart contracts and uses his/her private key $sk_{j}$ to decrypt $CT_j$:

\centerline{$ K \leftarrow UPPRE.Decrypt(sk_{j},CT_j)$}

2) The buyer downloads $M'$ from IPFS, and uses $K$ to decrypt $M'$:

\centerline{$M \leftarrow SE.Decrypt(M',K)$}

In the above fair trade protocol, it's necessary that the buyer trusts in the signature $Sig$ of the digital goods producer, which means that the correctness of the $IPFShash$ and $CT_i$ related to the digital goods $M$ is guaranteed by the digital goods producer. It is the only prerequisite for this fair trade protocol.

\textbf{Phase 0} to \textbf{phase 2} are regarded as preparation phases, and \textbf{Phase 8} is the decryption phase, so they are not included in the actual transaction process. In general, a transaction is established in \textbf{Phase 3}, and completed in \textbf{Phase 7}.

\subsection{The arbitration protocol}
In the fair protocol, a key step is that the seller computes the re-encryption key according to the request of the buyer. Therefore, for the verification of re-encryption key in \textbf{Phase 6}, if the verification fails, it means an error happens in the request $R$ or in the re-encryption key $rk_{ij}$.
In order to find out the dishonest party, it's necessary to introduce this arbitration protocol.

\textbf{Phase A1: Apply for arbitration.}

After submitting a request ${R=(\varphi,g_2,pk_{j1})}$, if the buyer does not receive the correct re-encryption key over a period of time, he/she submits a request to smart contracts for arbitration by disclosing the $sk_j$ used in \textbf{Phase 3}.

\textbf{Phase A2: Arbitration.}

When receiving a proof $sk_{j}=(x_{j1},x_{j2})$ from the buyer, smart contracts perform:

1) Check whether $pk_{j1}=g^{x_{j1}}$ and $g_2=g^{\frac{\varphi}{pk_{i1}^{x_{j1}}}}$ hold. If not, an error happens in the request $R$ from the buyer, so it's the buyer violating the fair protocol.

2) Otherwise, an error happens in the re-encryption key $rk_{ij}$ from the seller, so it's the seller violating the fair protocol.

Finally, smart contracts refund the payment to the buyer, and this transaction is terminated.

\subsection{Protocol analysis}

Case 1: The buyer is malicious, ignoring that the seller is honest or malicious.

If the buyer uses $x_{j1}$ that has been disclosed in the arbitration phase as part of the private key, and completes the purchase of digital goods, then the others can use this $x_{j1}$ to directly decrypt the corresponding $CT_j$ without purchasing digital goods. Therefore, smart contracts maintain a list $L_R$ to store accepted requests to prevent the buyer from using a repeated $(pk_j,sk_j)$ to purchase digital goods in \textbf{Phase 4}.

When the buyer is malicious, he/she may try to submit an incorrect $R$ and prevent the seller from calculating the corresponding re-encryption key $rk_{ij}$.
If the buyer wants the seller to suffer losses, the buyer must try to deceive smart contracts in the arbitration protocol. It means that the $x_{j1}$ submitted by the buyer must satisfy $pk_{j1}=g^{x_{j1}}$ and $g_2=g^{\frac{\varphi}{pk_{i1}^{x_{j1}}}}$. However, to pass these two verification equations, the buyer must correctly calculate $R$ by $UPPRE.Request()$ algorithm in \textbf{Phase 3}, which is inconsistent with the assumption.

Therefore, in this case, malicious buyer will be detected by the arbitration protocol.

Case 2: The buyer is honest and the seller is malicious.

The correctness of $Sig$ and smart contracts code can be verified by everyone, so the seller cannot engage in malicious behavior during the \textbf{Phase 2}. If the seller deliberately submits the wrong re-encryption key $rk_{ij}$, the smart contracts will detect that the $rk_{ij}$ is wrong through $UPPRE.VerifyRekey()$ algorithm in \textbf{Phase 6} and reject this submission. If the seller does not submit $rk_{ij}$ intentionally, the buyer can request for arbitration after the timeout.

In the arbitration protocol, the smart contracts will detect that the seller is malicious, then punish the seller and refund the buyer's prepayment.

Case 3: Both the buyer and the seller are honest.

If the buyer correctly calculates $R$ by $UPPRE.Request()$ algorithm and submits it to smart contracts, the seller can compute a valid re-encryption key $rk_{ij}$.
The smart contracts will verify the correctness of $rk_{ij}$ through $UPPRE.VerifyReKey()$ algorithm. Once the verification is passed, the transaction is successfully completed. It will convert the original ciphertext $CT_i$ into a transformed ciphertext $CT_j$ using the $UPPRE.ReEncrypt()$ algorithm, and transfer the owner of the prepayment from the buyer to the seller.
If both the buyer and the seller are honest, the seller will receive the payment released by the smart contracts. At the same time, the buyer will obtain the ability to decrypt $CT_j$ and get the corresponding symmetric key $K$. Obviously, in this case, the transaction will proceed smoothly and neither party will be punished or at a disadvantage.

Through the above analysis, we can draw the conclusion that the proposed protocol satisfies strong fairness.

\subsection{Protocol comparison}
The performance comparisons of some Blockchain-based fair protocols are summarized in Table \ref{tbl:ProAna}. Except \cite{corr/abs-2002-09689}, other protocols choose smart contracts as the Blockchain platform. Protocols in \cite{ZhaoYLHD19}\cite{XiongX19}\cite{DBLP:journals/access/HasanS18a}\cite{corr/abs-2002-09689} also need TTP to participate in the transactions.
Through analysis, except \cite{ZhaoYLHD19} and \cite{GuanSW18}, other protocols can achieve fairness. Only \cite{GuanSW18} and the proposed protocol support ciphertext publicity, while other protocols assume there is a secure channel to transfer the ciphertext. If digital goods can be sold repeatedly to different buyers without extra workload, we consider repeatable sale is supported. But only \cite{XiongX19}\cite{DBLP:journals/access/HasanS18a}\cite{GuanSW18} and the proposed protocol support this repeatable sale property. Interaction times of the proposed protocol is the least among all these protocols. Note that the interaction times in scheme \cite{GuanSW18} is related to $n$, where $n$ is the number of chunks that the plaintext of a digital good is split into, and $(1, n)$ represents the number from 1 to $n$.

\begin{table*}[]
\renewcommand\arraystretch{1.5}
\centering
\caption{Performances of several Blockchain-based fair protocols}
\begin{tabularx}{\linewidth}{XXXXXXXXXX}
\hline
\multicolumn{2}{l}{Schemes}                       &\cite{ZhaoYLHD19}&\cite{XiongX19}&\cite{DBLP:journals/access/HasanS18a}&\cite{DziembowskiEF18}&\cite{GuanSW18}&\cite{AsgaonkarK19}&\cite{corr/abs-2002-09689}& The Proposed Protocol\\ \hline
\multicolumn{2}{l}{Blockchain}                    & Smart contract  & Smart contract& smart contract &  Smart contract &Smart contract & Smart Contract&A public Blockchain  &Smart contract\\
\multicolumn{2}{l}{TPP}                           & Yes             & Yes           & Yes            & No             &  No           & No          & Yes        & No   \\
\multicolumn{2}{l}{Fairness}                      &  No             & Yes           & Yes            & Yes            &  No           & Yes         & Yes       &Yes   \\
\multicolumn{2}{l}{Ciphertext Publicity}          &  No             &$\backslash$   &$\backslash$    & No             &  Yes          & No          & No        &Yes   \\
\multicolumn{2}{l}{Repeatable Sale}               &  No             & Yes           &Yes             &No              &  Yes          & No          & No        &Yes   \\
\multirow{2}{*}{Interaction Times}& \qquad normal &     10          &     9         &     5          &     7          &  $2n+3$       & 6           &    7      & 7 or 6 \\
                                  & \qquad dispute&     11          &     9         &     5          &     7          &  $(1,2n+2)$   & 6           &    7      & 4     \\
                                  \hline
\end{tabularx}
\label{tbl:ProAna}
\end{table*}


\subsection{Experimental analysis}
The hardware configuration for testing keeps the same with Section IV. We implement the proposed protocol on smart contracts through Remix IDE and solidity programming language. The Remix IDE is an integrated development environment for the smart contracts. It provides basic functions such as compiling, deploying and executing contracts. The solidity version we use is 0.4.24, and we choose the JavaScript virtual machine as the environment for deploying and executing transactions. In this experiment, the symmetric encryption algorithm adopted is AES, and the symmetric key length is 256 bit. Two security parameters $l_q= 512$ and $l_q= 1024$ are used for UPPRE. The size of digital goods is 2MB.

By performing the entire fair trade protocol, the computation costs of buyers, sellers and smart contracts are analyzed.
Firstly, we focus on the time consumption for buyers and sellers.
The cryptographic algorithms for sellers include $SE.Encrypt()$, $UPPRE.Encrypt()$ and $UPPRE.ReKeyGen()$.
The cryptographic algorithms for buyers include $UPPRE.Request()$, $UPPRE.Decrypt()$ and $SE.Decrypt()$.
The time consumption of sellers and buyers is shown in \ref{tbl:time consumption}.
\begin{table}[H]
\renewcommand\arraystretch{1.5}
\centering
\caption{ Time consumption for sellers and buyers (ms)}
\begin{tabular}{p{1cm}p{3cm}p{1.5cm}p{1.5cm}}
\hline
Executor                & Algotirhms          &   $l_q= 512$        &  $l_q= 1024$ \\ \hline
\multirow{2}{*}{Sellers}&  $SE.Encrypt()$     &   29.5              &   29.8   \\
                        &  $UPPRE.Encrypt()$  &   1.9               &   5.3    \\
                        &  $UPPRE.ReKeyGen()$ &   0.7               &   2.9    \\
                        &   Total             &     32.0            &   38.0   \\
\multirow{2}{*}{Buyers} &  $UPPRE.Request()$  &   0.5               &   2.4    \\
                        &  $UPPRE.Decrypt()$  &   0.9               &   3.9    \\
                        &  $SE.Decrypt()$     &   16.7              &  17.0   \\
                        &   Total             &     18.1            &   23.3   \\

\hline
\end{tabular}
\label{tbl:time consumption}
\end{table}

Different from buyers and sellers, gas consumption is used to define the transaction cost on smart contracts, where gas is the currency unit to pay for transactions on the Ethereum Blockchain. The gas amount required for executing the smart contracts mainly depends the complexity of the program and the size of data. Transaction cost is defined to be the amount of gas consumed for the whole phases, and execution cost is the amount of gas for executing the codes, so execution cost is included in transaction cost.
We measure the gas consumption of the phases that related to smart contracts. Table \ref{tbl:GasCost1} and Table \ref{tbl:GasCost2} show the results on $l_q= 512$ and $l_q= 1024$ respectively.
\textbf{Phase 2} is used to deploy contracts, so sellers pay for the gas consumption. Generally speaking, one contract is deployed per digital goods.
\textbf{Phase 3} and \textbf{Phase 4} are used to make a request from buyers, so the buyers pay for the gas consumption.
\textbf{Phase 5} to \textbf{Phase 7} are used to generate re-encryption key by sellers, so the sellers pay for the gas consumption.
When there is a dispute, buyers apply for arbitration, so the gas consumption of \textbf{Phase A1} and \textbf{Phase A2} is paid by the buyer.

\begin{table}[H]
\renewcommand\arraystretch{1.5}
\centering
\caption{Gas consumption on smart  contracts with $l_q= 512$}
\begin{tabularx}{\linewidth}{XXXX}
\hline
Phases     &Payer & Transaction Cost (gas)  &  Execution  Cost (gas) \\ \hline
2          &Sellers   &   4971998               &  3816178    \\
3 and 4      &Buyers    &   415025                &  379353     \\
5, 6 and 7  &Sellers   &   1354995               &  1328795    \\
A1 and A2    &Buyers    &   536101                &  535229    \\
\hline
\end{tabularx}
\label{tbl:GasCost1}
\end{table}

\begin{table}[H]
\renewcommand\arraystretch{1.5}
\centering
\caption{Gas consumption on smart contracts with $l_q= 1024$}
\begin{tabularx}{\linewidth}{XXXX}
\hline
Phases     &Payer & Transaction Cost (gas)  &  Execution  Cost (gas) \\ \hline
2          &Sellers   &   5101896               &  3948036    \\
3 and 4      &Buyers    &   551325                &  502533     \\
5, 6 and 7  &Sellers   &   3865845               &  3835229    \\
A1 and A2    &Buyers    &   2043412               &  2033836    \\
\hline
\end{tabularx}
\label{tbl:GasCost2}
\end{table}

\section{Conclusion}
This paper propose to achieve the fairness of digital goods transactions on smart contracts  through a passive proxy re-encryption.
This scheme is backward-secure under collusion attacks, which means that the private key of delegator cannot be leaked, and the other ciphertext under this private key cannot be decrypted illegally.
Based on the proposed passive PRE scheme, a Blockchain based fair protocol for digital goods transactions is proposed. In this protocol, smart contracts can automatically transfer decryption right to the buyer after receiving his/her payment to ensure the fairness of transactions.

As PRE is a cryptographic primitive, although the bilinear maps are avoid, the computations over large number are still time-consuming. Especially for the smart contracts part, more gas consumption means higher transaction fees. If smart contracts are deployed on private Blockchain instead of Ethereum, there is no direct connection between transaction cost and currency, and the cost is not as high as it looks. Nevertheless, what we dedicate to are still efficient schemes or methods to achieve fair trade on Blockchain.


%



%
%

\ifCLASSOPTIONcaptionsoff
  \newpage
\fi



%



\bibliographystyle{unsrt}%
\bibliography{referencies}


%

%
%
%





\end{document}